\documentclass[11pt]{article}
\usepackage{times}
\usepackage{amsmath,amsthm,amssymb,setspace,enumitem,epsfig,titlesec,verbatim,color,array,eurosym,multirow}
\usepackage[sort&compress,comma,round,numbers]{natbib}
\usepackage[footnotesize,bf]{caption}
\usepackage[margin=2.5cm, includefoot, footskip=30pt]{geometry}
\usepackage{standalone}
\usepackage{tikz}
\usepackage{subcaption}
\usepackage{hyperref}
\usepackage[all]{hypcap}
\usepackage{tabularx}
\usepackage{booktabs}
\usepackage{blkarray}
\usepackage{pdfpages}
\usepackage[ruled,vlined]{algorithm2e}
\smallskip 

\def\alld{\texttt{ALLD}}
\def\tft{\texttt{TFT}}
\def\gtft{\texttt{GTFT}}
\def\rolemodel{\text{RM}}
\def\learner{\text{L}}

\def\strategy{s}
\def\esm{electronic supplementary material}

\newcommand{\FigBaseResults}{Figure~\ref{fig:expected_and_stochastic_for_donation}}
\newcommand{\FigDependenceParameters}{Figure~\ref{fig:cooperation_rate_over_benefit_and_beta}} 
\newcommand{\FigHigherMemory}{Figure~\ref{fig:cooperation_rate_all_updating_payoffs}}

\newcommand{\FigInvasionAnalysis}{{Supplementary Figure~3}} 
\newcommand{\FigMemoryOneParameters}{Supplementary Figure~7}

\titleformat{\section}{\sffamily \fontsize{12}{14}\bfseries}{\thesection}{1em}{}
\titleformat{\subsection}{\sffamily
\fontsize{11.5}{11.5}\bfseries}{\thesubsection}{1em}{}

\usepackage{tikz}
\usetikzlibrary{arrows}

\tikzset{treenode/.style = {align=center, inner sep=0pt, text centered,
  font=\sffamily}, arn_n/.style = {treenode, circle, white,
  font=\sffamily\bfseries, draw=black, inner sep=-6pt, fill=black, text
  width=1.5em},
  arn_r/.style = {treenode, circle, red, text width=1.5em, very thick, inner
    sep=4pt},
  arn_x/.style = {treenode, rectangle, draw=black, minimum width=0.5em, minimum
    height=0.5em}
}

\newtheoremstyle{plainCl1}
{9pt}
{9pt}
{\it}
{}
{\bfseries}
{.}
{0.2cm}
{}

\newtheoremstyle{plainCl2}
{9pt}
{9pt}
{\it}
{}
{\bfseries}
{$'$.}
{0.2cm}
{}

\theoremstyle{plainCl1}

\theoremstyle{plainCl2}

\title{\bf  \sffamily \Large Evolution of reciprocity 
with limited payoff memory\\}
\date{}
\author{\parbox[c]{16cm}{\centering \onehalfspacing 
Nikoleta E. Glynatsi$^1$,  Alex McAvoy$^{2,3, \dagger}$, Christian Hilbe$^{1, \dagger}$\\ \quad \\
$^{\rm 1}$Max Planck Research Group on the Dynamics of Social Behavior,\\ Max Planck Institute for Evolutionary Biology, Pl\"{o}n, Germany \\
$^{\rm 2}$School of Data Science and Society, University of North Carolina at Chapel Hill,\\ Chapel Hill, NC 27599 \\
$^{\rm 3}$Department of Mathematics, University of North Carolina at Chapel Hill,\\ Chapel Hill, NC 27599 \\
$^{\rm \dagger}$ A.M. and C.H. contributed equally to this work.}}

\begin{document}
\maketitle

\begin{abstract}
\noindent
Direct reciprocity is a mechanism for the evolution of cooperation in repeated social interactions. 
According to this literature, individuals naturally learn to adopt conditionally cooperative strategies if they have multiple encounters with their partner. 
Corresponding models have greatly facilitated our understanding of cooperation, yet they often make strong assumptions on how individuals remember and process payoff information. 
For example, when strategies are updated through social learning, it is commonly assumed that individuals compare their average payoffs.  
This would require them to compute (or remember) their payoffs against everyone else in the population.
To understand how more realistic constraints influence direct reciprocity, we consider the evolution of conditional behaviors when individuals learn based on more recent experiences.
Even in the most extreme case that they only take into account their very last interaction, we find that cooperation can still evolve. 
However, such individuals adopt less generous strategies, and they tend to cooperate less often than in the classical setup with average payoffs. 
Interestingly, once individuals remember the payoffs of two or three recent interactions, cooperation rates quickly approach the classical limit. 
These findings contribute to a literature that explores which kind of cognitive capabilities are required for reciprocal cooperation. 
While our results suggest that some rudimentary form of payoff memory is necessary, it already suffices to remember a few interactions.
\end{abstract}

~

\clearpage
\newpage


\section{Introduction}


Evolutionary game theory describes the dynamics of populations when an individual's fitness depends on the traits or strategies of other population members~\cite{hofbauer1998evolutionary, nowak:Nature:2004, hauert2005game,Traulsen:PhilTrans:2022}.  
This theory can be used to describe the dynamics of animal conflict~\citep{maynard-smith:Nature:1973}, cancer cells~\citep{Stein:PTRS:2023}, and of cooperation~\citep{nowak:Science:2006}. 
Respective models translate strategic interactions into games~\cite{smith1982evolution}. 
These games specify how individuals (players) interact, which strategies individuals can choose, and what fitness consequences (or payoffs) the different strategies have. 
In addition, these models also specify the mode by which successful strategies spread over time. 
In models of biological evolution, individuals with a high fitness produce more offspring; in models of cultural evolution, such individuals are imitated more often. 
Although biological and cultural evolution are sometimes treated as equivalent, there can be important differences~\citep{Wu2015,Smolla:PTRS:2021,Denton:TPB:2022}. 
For example, models of biological evolution do not require individuals to have any particular cognitive abilities.
Here, it is the evolutionary process itself that biases the population towards strategies with higher fitness. 
In contrast, in models of cultural evolution, individuals need to be aware of the different strategies present in the population, and they need to identify those strategies with a higher payoff. 
As a consequence, evolutionary outcomes may depend on how easily different behaviors can be learned~\citep{Chatterjee:JTB:2012}, and on how easy payoff comparisons are. 


These difficulties to learn strategies by social imitation are particularly pronounced in models of direct reciprocity. 
This literature follows Trivers' insight that individuals have more of an incentive to cooperate in social dilemmas when they interact repeatedly~\citep{trivers1971evolution}. 
In repeated interactions, individuals can condition their behavior on their past experiences with their interaction partner. 
They may use strategies such as Tit-for-Tat~\citep{rapoport:book:1965,axelrod1981evolution} or Generous Tit-for-Tat~\citep{molander:jcr:1985,Nowak1992tit} to preferentially cooperate with other cooperators. 
Such conditional strategies approximate human behavior fairly well~\citep{fischbacher:EconL:2001,Rand:TCS:2013,DalBo:AER:2019,Montero-Porras:SciRep:2022,Rossetti:ETH:2023} and they have also been documented in several other species~\citep{Carter:PRSB:2013,Schweinfurth:AnBehav:2019,Voelkl:PNAS:2015} -- although direct reciprocity is generally more difficult to demonstrate in animals~\citep{CluttonBrock:Nature:2009,Silk:CurrentBiology:2013,taborsky:CurrentBiology:2013}.
However, at the outset, it is not clear how easy it is to {\it learn} reciprocal strategies by social imitation. 
As one obstacle, even if others' strategies are perfectly observable, individuals might find it difficult to identify which ones have the highest payoff. 
After all, the payoff of a strategy of direct reciprocity is not determined by the outcome of any single round.
Rather, it is determined by how well this strategy fares over an entire sequence of rounds, against many different population members. 
In practice, such information might be both difficult to obtain and to process. 


Most models of direct reciprocity abstract from these difficulties~\citep{brauchli:JTB:1999,brandt:JTB:2006,ohtsuki:JTB:2007b,szolnoki:pre:2009b,imhof2010stochastic,van-segbroeck:prl:2012,grujic:jtb:2012,Martinez2012,stewart:pnas:2013,pinheiro:PLoSCB:2014,stewart:games:2015,Baek2016,McAvoy:ProcA:2019,glynatsi:SCR:2020,Schmid:PlosCB:2022,Murase:SciRep:2022,Cooney:BMB:2022,Chen:PNASNexus:2023}. 
They just assume individuals can easily copy the strategies of others. 
Similarly, they just assume that updating decisions are based on the strategies' average (or expected) payoffs, which are based on all rounds and all interactions. 
These assumptions create a curious inconsistency in how models represent an individual's cognitive abilities. 
On the one hand, when playing the game, individuals are often assumed to have restricted memory. 
Respective studies typically assume that individuals make their decisions each round based on the outcome of the last round only~\citep[with only a few exceptions, see Refs.][]{Hauert1997,van-veelen:PNAS:2012,Stewart2016,Li:NatCS:2022,Murase:PLoSCompBio:2023a}. 
Yet when learning new strategies, individuals are assumed to remember (or compute) each others' precise average payoff across many rounds and many interaction partners. 
Herein, we wish to explore whether this latter assumption is actually necessary for the evolution of reciprocity through social imitation. 
We ask whether individuals can learn to adopt reciprocal strategies even when learning is based on payoff information from a limited number of rounds.


To explore that question, we theoretically study imitation dynamics in the repeated prisoner's dilemma, using two extreme scenarios. 
The first scenario is the usual modeling approach. 
Here, individuals update their strategies based on their expected payoffs. 
We contrast this model with an alternative scenario where individuals update their strategies based on the very
last (one-shot) payoff they obtained. 
We find that individuals with limited payoff memory tend to adopt less generous strategies. 
Yet moderate levels of cooperation can still evolve. 
Moreover, as we increase the individuals' payoff memory to include the last two or three one-shot payoffs, cooperation rates quickly approach the rates observed in the classical baseline case. 


Overall, these findings suggest that while memory is important, already minimal payoff information may suffice for the evolution of direct reciprocity based on social learning. 
They also suggest that the classical model of reciprocity (based on expected payoffs) can often be interpreted as a useful approximation to more realistic models that include cognitive constraints.


\section{Model and Methods}\label{section:model}


To explore the impact of limited payoff memory, we adapt existing models of the evolution of direct reciprocity.
These models involve two different time scales. 
The short time scale describes the game dynamics. 
Here, individuals with fixed strategies are randomly matched to interact with each other in repeated social dilemmas. 
The long time scale describes the evolutionary dynamics. 
Here, individuals can update their repeated-game strategies based on the payoffs they yield. 
In the following, we introduce the basic setup of our model; all details and derivations are described in the \esm.\\


\noindent
{\bf Description of the game dynamics.} We consider a well-mixed population consisting of $N$~players.
Players are randomly matched in pairs to participate in a repeated donation game~\citep{sigmund2010calculus} with their respective co-player.
Each round, they can either cooperate (\(C\)) or defect (\(D\)). 
A cooperating player provides a benefit~\(b\) to the other player at their own cost~\(c\), with \(0 \!<\! c \!<\! b\). 
A defecting player provides no benefit and pays no cost. 
Thus, the players' payoffs in a single round are given by the matrix
\begin{align}
\bordermatrix{%
	& C & D \cr
C &\ b-c &\ -c\  \cr
D &\ b &\ 0\ \cr
} .
\end{align}
In particular, payoffs take the form of a prisoner's dilemma:
Mutual cooperation yields a better payoff than mutual defection ($b\!-\!c\!>\!0$), but each player individually prefers to defect independent of the co-player's action ($b\!>\!b\!-\!c$ and $0\!>\!-c$). 
To incorporate that individuals interact repeatedly, we assume that after each round, there is a constant continuation probability $\delta$ of interacting for another round. 
For $\delta\!=\!0$, we recover the case of a conventional (one-shot) prisoner's dilemma. 
Here, mutual defection is the only equilibrium. 
As $\delta$ increases, the game turns into a repeated game. Here, additional equilibria emerge, with some of them allowing for full cooperation~\citep{friedman:RES:1971,Akin:chapter:2016,hilbe:GEB:2015,stewart:pnas:2014}.


In a one-shot donation game, players can only choose among two pure strategies (they can either cooperate or defect).
In the repeated game, strategies become arbitrarily complex. 
Here, strategies are contingent rules, telling players what to do depending on the outcome of all previous rounds. 
For simplicity, in the following we assume individuals use {\it reactive strategies}~\citep{Nowak1992tit}. 
A reactive strategy only depends on the other player's action in the last round. 
Such strategies can be written as a three-dimensional tuple \(\strategy=(y, p,q)\).
The first entry \(y\) is the probability that the player opens with cooperation in the first round. 
The two other entries are the probabilities that the player cooperates in all subsequent rounds, depending on whether the co-player cooperated~($p$) or defected~($q$) in the previous round. 
The set of reactive strategies is simple enough to facilitate an explicit mathematical analysis~\citep{hofbauer1998evolutionary}. 
Yet it is rich enough to capture several important strategies of repeated games. 
For example, it contains \alld{} $=\!(0,0,0)$, the strategy that always defects. 
Similarly, it contains Tit-for-Tat, \tft{} $=\!(1,1,0)$, the strategy that copies the co-player's previous action (and that cooperates in the first round). 
Finally, it contains Generous Tit-for-Tat, \gtft $=\!(1,1,q)$, where $q\!>\!0$ reflects a player's generosity in response to a co-player's defection~\citep{molander:jcr:1985,Nowak1992tit}.  

In the short run, the players' strategies are taken to be fixed.
Players use their strategies to decide whether to cooperate in a series of repeated games against all other population members. 
In the long run, however, the players' strategies may change depending on the payoffs they yield, as we describe in the following.\\
 

\noindent
{\bf Description of the evolutionary dynamics.}
Herein, we assume population members update their strategies based on social learning. 
To model these strategy updates, we consider a pairwise comparison process~\citep{traulsen2007pairwise}. 
This process assumes that at regular time intervals, one population member is randomly selected, and given the chance to revise its strategy.
We refer to this player as the `learner'. 
With probability $\mu$ (reflecting a mutation rate), the learner simply adopts a random strategy (all reactive strategies have the same probability to be chosen). 
With the converse probability $1\!-\!\mu$, the learner randomly picks a `role model' from the population. 
The learner then compares its own payoff $\pi_\learner$ from the repeated game to the role model's payoff $\pi_\rolemodel$. 
The learner adopts the role model's strategy with a probability \(\varphi\) described by a Fermi function~\citep{blume:GEB:1995,szabo:PRE:1998}, 
\begin{equation} \label{Eq:rho}
    \varphi\left(\pi_\learner, \pi_\rolemodel\right) = \frac{1}{1\!+\! e^{- \!\beta\left(\pi_\rolemodel- \pi_\learner\right)}}.
\end{equation}
The selection strength parameter $\beta\!\ge\!0$ indicates how sensitive players are to payoff differences. 
For $\beta\!=\!0$, payoff differences are irrelevant, and the learner simply adopts the role model's strategy with probability one half. As the selection strength~$\beta$ increases, players are increasingly biased to imitate the role model only if it has the higher payoff. 


We deviate from previous models in how we interpret the payoffs $\pi_\learner$ and $\pi_\rolemodel$, which form the basis of the pairwise comparisons in Eq.~\eqref{Eq:rho}. 
In previous work, these payoffs are taken to be the respective players' expected payoffs. 
We interpret that setup as a model with perfect payoff memory. 
There, the payoffs  $\pi_\learner$ and $\pi_\rolemodel$ represent an average over all possible repeated games the two individuals have played with all population members (\FigBaseResults, upper left panel). 
The use of expected payoffs is mathematically convenient, because explicit formulas for these payoffs are available~\citep{hofbauer1998evolutionary}.
Herein, we compare this model of perfect payoff memory to a model with limited payoff memory. 
In that model, the players' payoffs $\pi_\learner$ and $\pi_\rolemodel$ are taken to be the payoffs that each player received in their very last round prior to making social comparisons. 
That is, players only consider the very last repeated game they participated in, and there they only take into account the outcome of the very last round (\FigBaseResults, lower left panel). 
This assumption could reflect, for example, a strong recency bias in how individuals evaluate payoffs.  
In addition to this extreme case of limited payoff memory, later on we also explore cases in which players take into account the outcome of two, three, four, or more recent rounds. 


Both in the case of perfect and limited memory, we iterate the elementary strategy updating step described above for many time steps. 
This gives rise to a stochastic process that describes which strategies players adopt over time. 
We explore the dynamics of this process mathematically and with computer simulations.
For the results presented in the following, we assume that mutations are rare (\(\mu\!\rightarrow\! 0\)). 
This assumption is fairly common in evolutionary game theory, because it makes some computations more efficient~\citep{fudenberg:JET:2006,wu:JMB:2012,mcavoy:jet:2015}, and because the results can be interpreted more easily.  
However, in Section~3 of the \esm{} we show that our main results continue to hold for strictly positive mutation rates.


\section{Results}


\noindent
{\bf Stability of cooperative populations.}
To get some intuition for the differences between perfect and limited payoff memory, we first analyze when cooperation is stable in either scenario.
To this end, we consider a resident population in which all players but one adopt a strategy of Generous Tit-for-Tat, \gtft{} $=\!(1,1,q)$. 
The remaining mutant player adopts \alld. 
We say {\it cooperation is stochastically stable} if the single mutant is more likely to imitate the residents than vice versa. 
For simplicity, we consider a large population~($N\!\rightarrow\!\infty$) and strong selection~($\beta\!\rightarrow\!\infty$).
More general results are derived in the \esm. 


In the case of perfect payoff memory, it is straightforward to characterize when cooperation is stochastically stable. 
Here, we simply need to compute the players' expected payoffs. 
Because the population mostly consists of residents, and because residents mutually cooperate with each other, their expected payoff is $\pi_\gtft = b\!-\!c$. 
On the other hand, the defecting mutant only interacts with residents. 
Given the residents' strategy, the mutant receives a benefit in the first round, and in every subsequent round with probability $q$. 
As a result, the mutant's expected payoff is $\pi_\alld \!=\! (1\!-\!\delta\!+\!\delta q)b$. 
For perfect payoff memory, the requirement for cooperation to be stochastically stable reduces to the condition $\pi_\gtft > \pi_\alld$. 
This yields
\begin{equation} \label{Eq:PerfectMemory}
q < 1\!-\!\frac{c}{\delta  b}.
\end{equation}
In particular, we recover the previous observation that $q\!=\!1-c/(\delta b)$ is the maximum generosity that cooperators should have~\citep{molander:jcr:1985,Nowak1992tit,Schmid:NHB:2021}. 
Because $q\!\ge\!0$, we also conclude that cooperation can only be stable if $\delta \!>\! c/b$.
Again, this condition for the feasibility of direct reciprocity is the condition found in the literature~\citep{nowak:Science:2006}.


The logic of the case with limited payoff memory is somewhat different. 
Here we need to compute how likely each player obtains one of the four possible payoffs $\{b\!-\!c, -c, b, 0\}$ in the very last round of a game, before they make social comparisons.
Because residents almost always interact with other residents, their last one-shot payoff is $\pi_\gtft = b\!-\!c$ almost surely. 
For the defecting mutant, there are two possibilities. 
({\it i})~ If the mutant's co-player happens to cooperate in the last round, the mutant receives $\pi_\alld\!=\!b$.
This case occurs with probability $1\!-\!\delta\!+\!\delta q$. 
({\it ii})~If the co-player defects in the last round, the mutant receives $\pi_\alld=\!0$. 
This occurs with the converse probability $\delta(1\!-\!q)$.
Because $b\!-\!c\!<\!b$, residents tend to imitate the mutant in the first case. 
Because $b\!-\!c\!>\!0$, mutants tend to imitate the resident in the second case. 
Cooperation is stochastically stable if the first case is less likely than the second. 
This yields the condition
\begin{equation} \label{Eq:LimitedMemory}
q < 1\!-\!\frac{1}{2 \delta}.
\end{equation}
Interestingly, this condition no longer depends on the exact payoff values $c$ and $b$. 
This independence arises because of our assumption of strong selection, in which case only the payoff ordering $b\!>\!c\!>\!0$ matters. Because $q$ is non-negative, condition~\eqref{Eq:LimitedMemory} can only be satisfied if $\delta\!>\!1/2$. That is, players need to interact in more than two rounds in expectation. 


By comparing the two cases, we find that payoff memory affects whether a conditionally cooperative strategy ($1,1,q$) is viable. 
With perfect memory, the maximum generosity $q$ needs to satisfy Eq.~\eqref{Eq:PerfectMemory}.
In particular, this generosity can become arbitrarily large, provided the game's benefit-to-cost ratio~$b/c$ and the continuation probability~$\delta$ are sufficiently large. 
In contrast, with limited payoff memory, the maximum generosity is bounded by one half, and it is independent of the benefit-to-cost ratio.\\


\noindent
{\bf Evolutionary dynamics of reciprocity.}
To explore whether the previous static observations describe the dynamics of evolving populations, we turn to simulations.
We have run separate simulations for perfect and limited payoff memory. 
In each case, we consider both a low and a high benefit of cooperation ($b/c\!=\!3$ and $b/c\!=\!10$, respectively).  
For each simulation, we record which strategies ($y,p,q$) the players adopt over time.
\FigBaseResults{} depicts the conditional cooperation probabilities $p$ and $q$ (we omit the opening move \(y\) because we use a discount factor~\(\delta\) close to one, such that first-round behavior is largely irrelevant). 
In all simulations, we find that the players' strategies cluster in two regions of the strategy space. 
The first region corresponds to a neighborhood of \alld{} with $(p,q)\!\approx\!(0,0)$.
The second region corresponds to a thin strip of cooperative strategies with $(p,q)\!\approx\!(1,q)$. 
Within this strip, we observe that most strategies satisfy the constraints on $q$ imposed by the inequalities~\eqref{Eq:PerfectMemory} and~\eqref{Eq:LimitedMemory}. 
That is, with perfect memory, most evolving strategies have $q\!<\!1\!-\!c/b$, whereas with limited payoff memory, most strategies have $q\!<\!1/2$. 
In particular, for limited payoff memory, changes in the benefit parameter have no effect on the qualitative distribution of strategies. 


In each case, the evolutionary dynamics follow a similar cyclic pattern~\citep[as described in Refs.][]{imhof2010stochastic, Nowak1992tit}:
Resident populations of defectors are most likely invaded by strategies close to \tft. 
Once the population adopts conditionally cooperative strategies  $(1,1,q)$, neutral drift may introduce larger values of generosity~$q$. 
If the resident's generosity~$q$ violates the conditions~\eqref{Eq:PerfectMemory} and~\eqref{Eq:LimitedMemory}, defectors can re-invade, and the cycle starts again. 
The relative time spent near \alld{} and near the strip of conditionally cooperative strategies depends on the considered memory setting (\FigInvasionAnalysis, depicting the case of high benefits). 
For perfect memory, we find that \alld{} is replaced relatively quickly by more cooperative strategies. 
Here, it takes on average 159 invading mutants until \alld{} is successfully replaced. 
In contrast, for limited memory, \alld{} is more robust, resisting on average 798 mutant strategies. 
This picture reverses when we consider an initial population that adopts \gtft. 
Such populations are much more robust under perfect memory than they are under limited memory. 
Overall, we find that the impact of memory on the population's average cooperation rate is substantial. 
For perfect memory, this rate is 52\% for low benefits, and 98\% for high benefits. 
For limited payoff memory, the evolving cooperation rates are smaller but still strictly positive, with 37\% cooperation for low benefits and 51\% cooperation for high benefits (\FigBaseResults). 


To further investigate the influence of different parameters, we have systematically varied the benefit~$b$ and the selection strength~$\beta$ in~\FigDependenceParameters.
According to \FigDependenceParameters{\it a}, perfect memory consistently results in a higher cooperation rate, and this relative advantage further increases with an increasing benefit~$b$. 
Interestingly, for limited payoff memory, the cooperation rate remains stable at approximately 50\% once \(b \!\ge\! 5\).
This again reflects our earlier observation that the feasibility of cooperation in this scenario is largely independent of the exact values of $b$ and $c$, as described by Eq.~\eqref{Eq:LimitedMemory}. 
With respect to the effect of different selection strengths, \FigDependenceParameters{\it b} suggests that both perfect and limited payoff memory yields similar cooperation rates for weak selection \(\beta \!<\! 1\). 
Beyond weak selection, increasing selection has a positive effect under perfect payoff memory, but a negative effect under limited payoff memory.\\


\noindent
{\bf Beyond reactive strategies.} While the results presented in the main text focus on reactive strategies, the patterns we observe do not seem to depend on the considered strategy space. 
To illustrate this point in more detail, in the \esm{} we consider the dynamics among memory-1 strategies. 
Here, players take into account both their co-player's and their own last move, see Refs.~\citep{nowak:Nature:1993,imhof:JTB:2007}. 
Also in that case, we observe that perfect memory leads to systematically higher cooperation rates (Supplementary Figures~5,6).
Again, this advantage of perfect memory is particularly pronounced for strong selection, or when there is a high benefit of cooperation (\FigMemoryOneParameters).\\ 


\noindent 
{\bf The effect of increasing individual payoff memory.}
So far, we have taken a rather extreme interpretation of limited payoff memory. 
In the respective scenario, we assumed that individuals update their strategies based on their experience in a single round of the prisoner's dilemma, against a single co-player. 
The limited payoff memory framework can be expanded in various ways. 
In particular, individuals may recall a larger number of rounds, they may recall their interactions with several co-players, or both. 
To gain further insights on the impact of payoff memory, we explore four additional scenarios. 
In the first scenario, players recall the payoffs they obtained in the last two rounds against a single co-player. 
In the second scenario, players recall their last-round payoffs against two co-players. 
In the third scenario, they recall the two last rounds against two co-players. 
Finally, in the last scenario, players update based on the average payoff they
receive over all rounds with a single co-player (further extensions are possible, but we do not explore them here).


For most scenarios, we can again derive an analytical condition for when cooperation is stochastically stable. 
As before, we assume populations are large and that selection is strong. 
For simplicity, we also assume that the game continues almost certainly after each round (i.e., $\delta$ approaches one). 
The details of this analysis can be found in the \esm. 
In the first two scenarios, we interestingly find that for $b\!>\!2c$, cooperation is stochastically stable when $q\!<\! \frac{\sqrt{2}}{2}\approx 0.707$. 
Comparing this condition with the more stringent condition in Eq.~\eqref{Eq:LimitedMemory} suggests that there are now more conditionally cooperative strategies that can sustain cooperation. 
Hence, cooperation should evolve more easily.
In the last scenario, we find that cooperation is stochastically stable when \(q
< 1 - \frac{c}{b}\), which is the same condition as in
Eq.~\eqref{Eq:PerfectMemory}, even though only a single co-player is considered
instead of the whole population. 


We complement these analytical results with additional simulations, see \FigHigherMemory.
We observe that a minimal increase in the players' payoff memory (compared to the baseline case with a single round recalled) can promote cooperation considerably. 
Specifically, in all four scenarios with extended memory, we see similar cooperation rates, and they approach the rates observed under perfect memory. 
These results suggest that while it takes {\it some} payoff memory to sustain substantial cooperation rates, the requirements on memory seem to be rather modest. 
Already remembering a few interactions, either with the same co-player or across different co-players, may provide players with enough information to adopt reciprocal strategies.


\section{Discussion}\label{section:discussion}
In economics, if payoff is measured in terms of money and a decision is to be made at time $t$ in the future, then currency accumulated early on is weighted more in that decision because it has more time to accumulate interest. Such a model discounts the future relative to the past. It also does not necessarily require ``memory'' because rewards are accumulated into a factor used in decision-making; the specific time stamps of rewards do not themselves provide better information beyond their effects on total payoff. In a similar fashion, the probabilistic interpretation of discounting as a continuation probability \citep{axelrod1981evolution} also effectively discounts the future relative to the past. A foraging animal deciding between two behaviors might tend to choose the one that yields a moderate reward sooner relative to a larger reward later, since earlier rewards (e.g., food) contribute to immediate survival, and there is no guarantee that later rewards will happen at all \citep{stephens1986foraging}.

Within the context of a single repeated game, the model we consider here is, in some ways, dual to the classical model of temporally-discounted rewards in repeated games. Instead of making decisions based on expected rewards in the future, we consider individuals who make decisions based on actual rewards in the past. Thus, the ability to estimate the future payoff of a strategy is replaced by the memory of how this strategy previously fared against others. This involves two time scales: interaction partners and rounds within those interactions. As a result, we are dealing with a model that discounts the past rather than the future. Intriguingly, treating payoffs in this manner is reminiscent of the reward-smoothing technique of ``eligibility traces'' in reinforcement learning \citep{sutton:MIT:2018}, which uses past rewards (discounted appropriately) to shape present perceived payoff. There is a sound basis for this method in neuroscience, where rewards and (temporal-difference) learning are associated to dopaminergic neurons \citep{schultz:JN:1998} and spike-timing-dependent plasticity \citep{dan:Neuron:2004}. This suggests a more biologically-encoded interpretation of memory, which is equally applicable to models of direct reciprocity where rewards have a neurological basis.

Of course, the precise nature of ``memory'' also depends on what payoffs in a game represent, which should be taken into account when applying game-theoretic models. For example, a payoff stream of monetary currency might truly accumulate and not require memory on the parts of agents. Even in the context of money, however, not all of what was obtained in the past is necessarily available at the time a decision is made, which brings memory into play. The serial position effect in human psychology shows that in an ordered list of items (e.g., words), humans tend to have difficulty remembering the entirety of sequences, demonstrating moderate recall for those items coming earlier (primacy effect), substantial recall for those coming later (recency effect), and lower recall for those in between \citep{murdock:JEP:1962}. It is therefore reasonable that when presented with a stream of payoffs, whether on the timescale of pairing for repeated interactions or in a stream of one-shot games, players might be able to effectively incorporate only the most recent payoffs.

In fact, even beyond specific psychological considerations, a curious interpretation of payoffs arises from the formula commonly used for expected payoffs in repeated games. If $\delta\in\left[0,1\right)$ is the probability of continuing to another round in the game, then the expected payoff to an agent is $\left(1-\delta\right)\sum_{t=0}^{\infty}\delta^{t}u_{t}$, where $u_{t}$ is the reward the agent receives in the stage game at time $t$. Here, additional stochasticity arises due to uncertainty in the game length, and an agent might not be able to compute his or her expected payoff for use in decision-making. The probability that the game terminates after the interaction at time $T$ is $\delta^{T}\left(1-\delta\right)$, in which case $\left(1-\delta\right)\sum_{t=0}^{\infty}\delta^{t}u_{t}$ is exactly the expected payoff the agent receives at time $T$, i.e. \emph{in the last round}. As an unbiased estimator of this expectation, the agent might thus use $u_{T}$ as a proxy for ``success'' when evaluating his or her behavior. This gives a purely model-driven justification for why considering payoff in the last round of the game can result in more realistic extensions of traditional models.

We note that the expected \emph{total} payoff in the game, $u_{0}+u_{1}+\cdots +u_{T}$, is given by $\sum_{t=0}^{\infty}\delta^{t}u_{t}$. This version of ``expected payoff'' appears less common in the literature on direct reciprocity than its normalization, $\left(1-\delta\right)\sum_{t=0}^{\infty}\delta^{t}u_{t}$, likely owing to the fact that payoffs can grow arbitrarily large with sufficiently long time horizons ($\delta\rightarrow 1^{-}$). Non-normalized payoffs interfere with selection intensity ($\beta$) in models of social imitation, which is (presumably) why they appear less frequently in the literature. On that point, we note that many of the differences between realized and expected payoffs disappear in the limit of weak selection \citep{mcavoy:NHB:2020}. Non-weak selection can introduce substantial differences between models with realized and expected payoffs \citep{mcavoy:PLOSCB:2021}, which is especially important to understand in models of social systems with cultural transmission \citep{cavalli1981cultural}.

Our main contribution is an application of these ideas to direct reciprocity, which is one of the key mechanisms to explain why unrelated individuals might cooperate~\citep{nowak:Science:2006}. 
According to this mechanism, cooperation pays if it makes the interaction partner more cooperative in future. 
To describe which strategies are most effective, the previous theoretical literature assumes that the evolutionary dynamics are driven by the players' expected payoffs~\citep{brauchli:JTB:1999,brandt:JTB:2006,ohtsuki:JTB:2007b,szolnoki:pre:2009b,imhof2010stochastic,van-segbroeck:prl:2012,grujic:jtb:2012,Martinez2012,stewart:pnas:2013,pinheiro:PLoSCB:2014,stewart:games:2015,Baek2016,McAvoy:ProcA:2019,glynatsi:SCR:2020,Schmid:PlosCB:2022,Murase:SciRep:2022}.
To the extent that strategies are learned (not inherited), this assumption seems to impose rather stringent requirements on the individuals' cognitive abilities. 
In the most extreme case, this would require individuals to remember (or compute) their payoffs against all population members, for all possible ways in which their repeated games may unfold. 
This assumption introduces a curious inconsistency in how these models represent an individual's cognitive abilities. 
For playing their games, individuals are often assumed to only recall the outcome of the very last round. 
Yet to update their strategies, individuals are implicitly assumed to have a record of the outcome of all rounds, across all interaction partners. 

It is natural to ask, then, to what extent perfect payoff memory is in fact required for the evolution of reciprocity. 
To this end, we consider a model in which individuals only remember the payoff of their very last interaction, or the payoffs of the last few interactions. 
By only considering an individual's most recent experiences, the evolutionary process is subject to additional stochasticity. 
Strategies that perform well on average (across an entire repeated game and across many interaction partners) may still get replaced if the respective player happened to yield an inferior payoff in the very last round. 
A similar element of stochasticity has been previously explored in the context of one-shot (non-repeated) games~\citep{sanchez:JTB:2005,roca:PhysicalReview:2006,Traulsen:JTB:2007,Woelfing:JTB:2009,Hauert:PRE:2018}. 
This literature studies which strategies are selected for when individuals only interact with a finite sample of population members. 
In the respective models, individuals can only choose among two strategies. 
They can either cooperate or defect, and stochastic sampling affects which of these two strategies is favored. 
Instead, in repeated games, players have access to a large set of strategies~\citep[in our case, all reactive strategies;][]{nowak:APC:1989}. 
Here, stochastic sampling does not only affect whether cooperative or non-cooperative strategies are favored; it also affects {\it which} conditionally cooperative strategies are favored. 

To address these questions, we combine analytical methods and computer simulations. 
In the most extreme case, we consider individuals who update their strategies based on only one piece of information:  the last round of a single repeated game. 
For that case, we find that individuals are less generous, and they tend to be less cooperative overall~(\FigBaseResults). 
However, once individuals update their strategies based on two or more recent experiences, overall cooperation rates quickly approach the levels observed under perfect payoff memory~(\FigHigherMemory). 
These findings suggest that models based on expected payoffs can serve as a useful approximation to more realistic models with limited payoff memory. 
Our findings also contribute to a wider literature that explores which kinds of cognitive capacities are required for reciprocal altruism to be feasible~\citep[e.g.,][]{Stevens:fip:2011,Volstorf:PlosOne:2011}. 
While more payoff memory is always favorable, reciprocal cooperation can already be sustained if individuals have a record of two or three past outcomes. We believe that this kind of result, derived entirely within a theoretical model, is crucial for making model-informed deductions about reciprocity in natural systems.\\[0.5cm]



\noindent
{\bf Data accessibility.}
All data and code used in this manuscript are openly available at\newline
\texttt{https://zenodo.org/records/10066227}.\\






{
{\setlength{\bibsep}{0\baselineskip}
\bibliographystyle{naturemag}
\bibliography{bibliography}
}


\clearpage
\newpage

\begin{figure}[t]
    \centering
    \includegraphics[width=\textwidth]{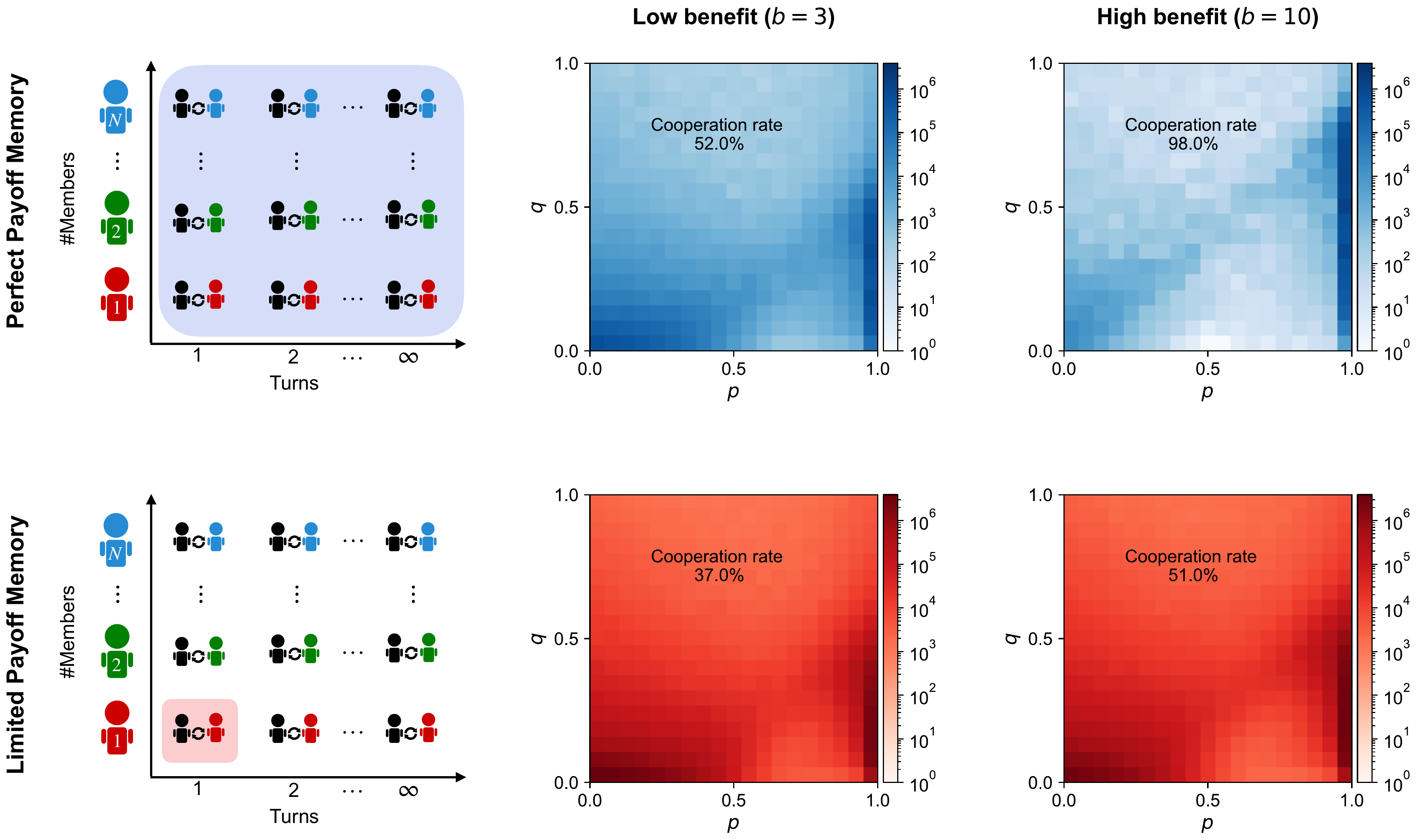}
    \caption{{\bf Evolutionary dynamics under perfect and limited payoff memory.}
    The leftmost panels give a schematic overview of the two main scenarios we compare. 
    The two scenarios differ in how many past interactions individuals take into account when updating their strategy. 
    In the scenario with perfect payoff memory, individuals consider all their past interactions (against all population members, and taking every turn of each repeated game into account). 
    In the scenario with limited payoff memory, individuals only consider their very last interaction (against one specific population member, taking into account only one round of the repeated game). 
    The four panels  on the right side depict the outcome of evolutionary simulations for repeated games with either a low or a high benefit of cooperation. 
    Colors represent how often the respective region of the strategy space is visited over time. 
    In all four panels, two regions are visited particularly often. 
    One region corresponds to a neighborhood of \alld{} with $p\approx q\!\approx\!0$ (lower left corner). 
    The other region corresponds to a strip of conditionally cooperative strategies with $p\!\approx\! 1$ and $q$ satisfying the constraints \eqref{Eq:PerfectMemory} and \eqref{Eq:LimitedMemory}, respectively (lower right corner). 
    The resulting average cooperation rate depends on which of these two neighborhoods is visited more often. 
    Simulations are run for $T\!=\!10^7$ time steps, using a cost $c\!=\!1$, a continuation probability of $\delta\!=\!0.999$ and a selection strength of $\beta\!=\!1$, in a population of size $N\!=\!100$.}
\label{fig:expected_and_stochastic_for_donation}
\end{figure}

\clearpage
\newpage

\begin{figure}[t]
    \centering
    \includegraphics[width=.75\textwidth]{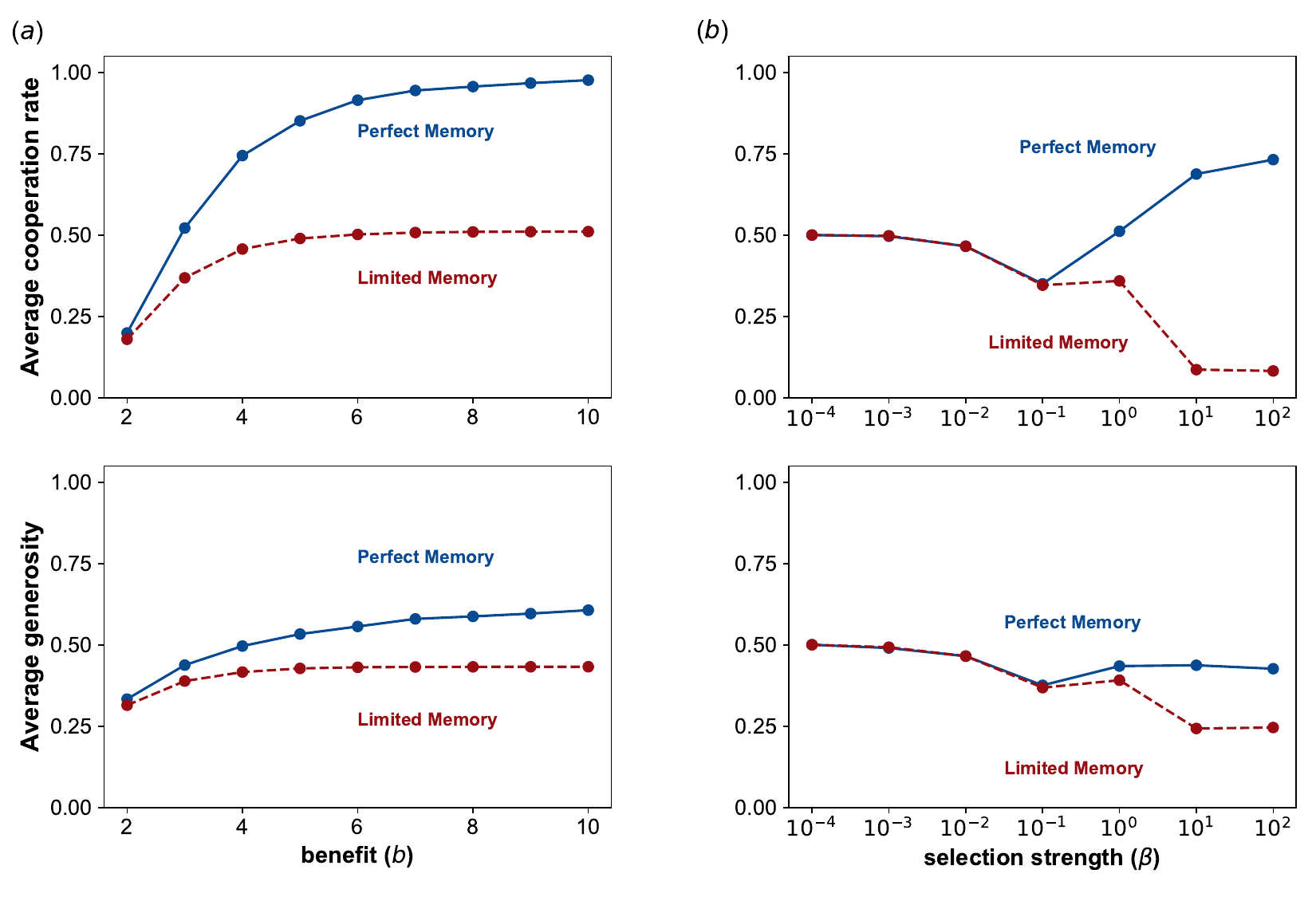}
    \caption{{\bf Evolution of direct reciprocity for different parameter values.} 
    To explore the robustness of our results, we have run simulations for different benefit values (left panels, {\it a}), and for different selection strengths (right panels, {\it b}). 
    In each case, we record the resulting average cooperation rate over the entire simulation (upper panels). 
    In addition we record the individuals' average generosity. 
    Here, we only take into account those residents with $p\! \approx\! 1$ and we compute
    the average of their cooperation probability~$q$. 
    These simulations suggest that perfect payoff memory consistently leads to more cooperation and more generosity. 
    Unless explicitly varied, the parameters of the simulation are $N\!=\!100$, $b\!=\!3$, $c\!=\!1$, $\beta\!=\!1$, $\delta\!=\!0.99$. 
    Simulations are run for $T\!=\!5\times 10^7$ time steps for each parameter combination.}
    \label{fig:cooperation_rate_over_benefit_and_beta}
\end{figure}

\clearpage
\newpage

\begin{figure}[t]
  \centering
  \includegraphics[width=\textwidth]{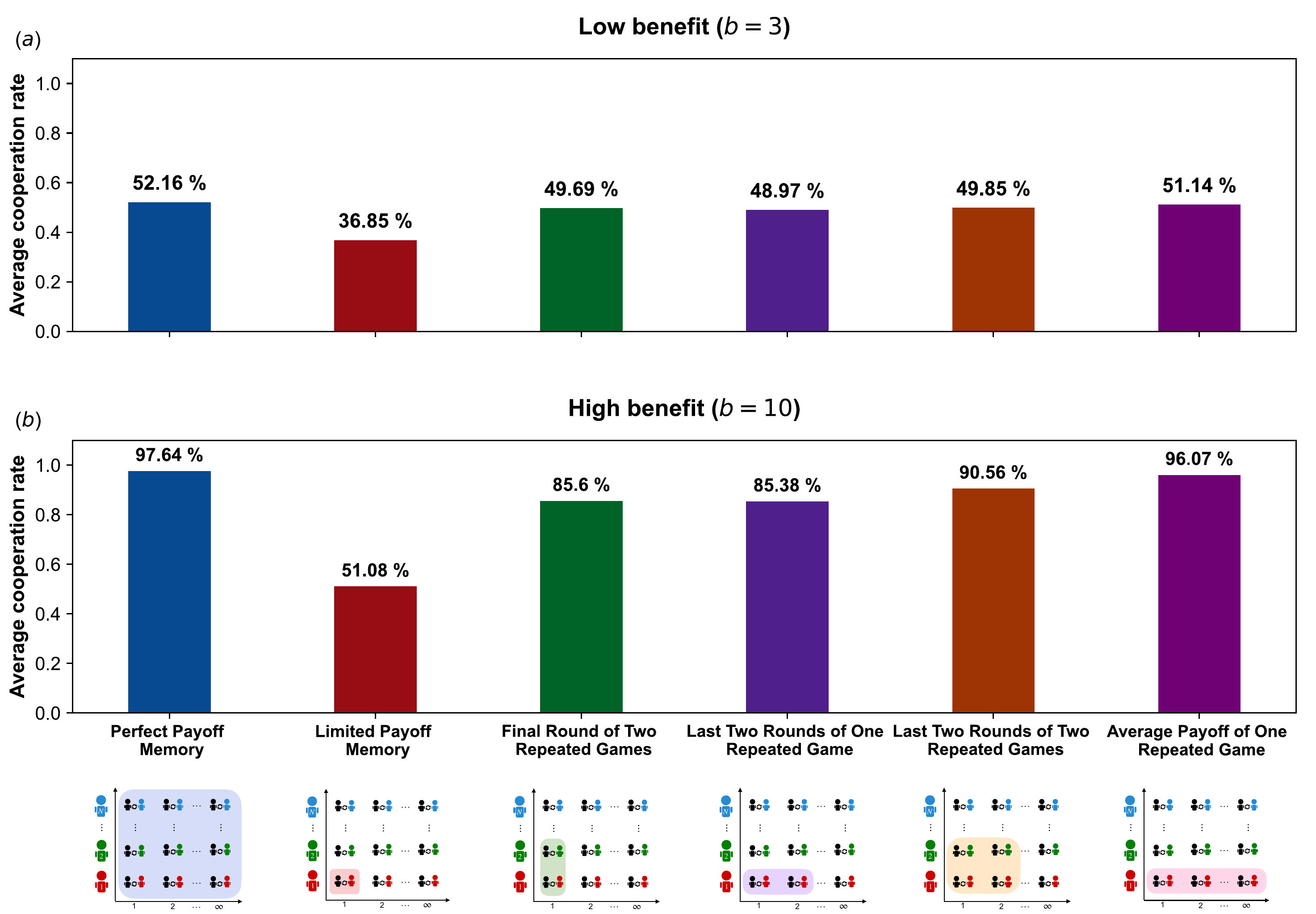}
  \caption{{\bf Average cooperation rates for different payoff memories.}
  We vary how much information individuals take into account when updating their strategies. 
  From left to right, we consider the following cases.
  ({\it i}) Updating occurs based on expected payoffs (perfect memory), 
  ({\it ii}) it occurs based on the last round of one interaction (limited memory), 
  ({\it iii}) based on the last round of two interactions,
  ({\it iv}) based on the last two rounds of one interaction,
  ({\it v}) based on the last two rounds of two interactions, and
  ({\it vi}) based on the average payoff of one interaction.
  Again, simulations are run either for a comparably low benefit of cooperation~($b/c\!=\!3$), or for a high benefit~($b/c\!=\!10$). 
  We observe that perfect memory always yields the highest cooperation rate. 
  However, when individuals take into account at least two past interactions -- cases ({\it iii}) to ({\it vi}) -- evolving cooperation rates are close to this optimum. 
  Baseline parameters are the same as in \FigDependenceParameters.}
\label{fig:cooperation_rate_all_updating_payoffs}
\end{figure}

\clearpage
\newpage



\section*{Supplementary material}

# Electronic supplementary material
# Evolution of reciprocity with limited payoff memory

Nikoleta E. Glynatsi, Alex McAvoy, Christian Hilbe

This document provides further details on our methods and derivations, and it contains additional simulation results. Section 1 summarizes the model. In particular, we provide further details on our implementation of the evolutionary dynamics, and our use of the rare-mutation limit. In Section 2, we derive analytical results for the various settings we consider. These settings differ in what kind of payoff information individuals take into account when updating their strategies. In the perfect-memory setting, individuals take into account all their interactions against all co-players. In the limited-memory setting, they only consider the very last round of their very last interaction. In addition, we describe several model extensions in which the amount of information taken into account is in between these two extremes. Finally, Section 3 presents further simulation results. In particular, we confirm that our main results continue to hold (*i*) when mutations are no longer rare, and (*ii*) when players use memory-one strategies instead of reactive strategies.

## 1 Description of the model

**Summary of the model.** As described in the main text, we study cooperative behavior in a population of size $N$, with $N$ being even. The dynamics unfold on two time scales. The short time scale describes the game dynamics. Here the $N$ individuals are randomly matched to form $N/2$ pairs to interact in a repeated prisoner's dilemma. Each round, individuals can choose whether to cooperate ($C$) or defect ($D$). In the most general setting, the resulting one-shot payoffs can be summarized by the payoff matrix

$$\begin{array}{c} \\ C \\ D \end{array} \begin{pmatrix} C & D \\ R & S \\ T & P \end{pmatrix}. \qquad (1)$$

Here, $R$ is interpreted as the reward for mutual cooperation, $S$ is the sucker's payoff, $T$ is the temptation, and $P$ is the punishment payoff [1]. Throughout this work, we parametrize these payoffs as $R=b-c$, $S=-c$, $T=b$, and $P=0$, where $b$ and $c$ are the benefit and cost of cooperation, respectively, with $b>c>0$. After each round, players learn their co-player's previous action. Then the game continues for another round with

probability $\delta$. Players make their decisions whether to cooperate in any given round based on their reactive strategies $s = (y, p, q)$. The entry $y$ determines a player's first-round cooperation probability. The other entries $p$ and $q$ determine the player's cooperation probability in all subsequent rounds. The probability is $p$ if the co-player cooperated in the previous round, and it is $q$ if the co-player defected. On this short time scale, the players' strategies are fixed, and players are consecutively matched to play repeated games with all other population members.

The long time scale describes the evolutionary dynamics. Here, players are allowed to update their strategies based on the payoffs they yield. We model these strategy updates with a pairwise comparison process [2]. This process assumes that at regular time intervals, one player is randomly selected from the population. We refer to this player as the 'learner' (L). The learner is given an opportunity to update its strategy. There are two possibilities for how this update may occur. With probability $\mu$, the player's strategy mutates randomly. In that case, the player's new strategy is drawn uniformly from the space of all reactive strategies $[0,1]^3$. With probability $1-\mu$, the player compares itself to another population member. To this end, the player randomly picks another individual from the population (referred to as the 'role model', RM). The learner adopts the role model's strategy with a probability $\varphi$, given by

$$\varphi(\pi_\text{L}, \pi_\text{RM}) = \frac{1}{1+e^{-\beta(\pi_\text{RM}-\pi_\text{L})}}. \tag{2}$$

The parameter $\beta \geq 0$ is the selection strength. It determines how important payoff differences are for the learner's decision to imitate the role model. The variables $\pi_\text{RM}$ and $\pi_\text{L}$ refer to the payoffs of the role model and the learner, respectively. The exact value of these payoffs depend on the players' memory. We say players have *perfect memory* when $\pi_\text{RM}$ and $\pi_\text{L}$ are given by the players' expected payoffs (across all rounds and across all possible co-players). We say players have *limited memory* when $\pi_\text{RM}$ and $\pi_\text{L}$ are given by the players' realized payoff in the very last round of the game with their very last interaction partner. In addition, we consider several model extensions in which individuals have memory capacities in between these two extremes. We provide a detailed description of these different settings and the resulting payoffs in Section 2.

**Evolutionary simulations for the rare-mutation limit.** To simulate the evolutionary dynamics of the pairwise comparison process, it is sometimes useful to assume that mutations are rare, $\mu \to 0$. In that case, whenever a mutant strategy appears, it either fixes in the population or goes extinct before the next mutant appears. As a result, at any given time there are at most two different strategies present in the population [3–5]. This assumptions makes computations more efficient, and it makes some of the results easier to interpret. In the following, we describe our implementation of the process in the rare-mutation limit in more detail.

Initially, the process starts with a population where all members use the same strategy (referred to as the resident strategy, R). Then one individual adopts a mutant strategy selected uniformly at random from the set

**Algorithm 1:** Evolutionary process in the limit of rare mutations
- $N \leftarrow$ population size;
- resident $\leftarrow$ starting resident;
- **while** $t <$ *maximum number of steps* **do**
  - mutant $\leftarrow$ random strategy;
  - fixation probability $\leftarrow \rho_M$;
  - **if** $\rho_M >$ *random: $i \to [0, 1]$* **then**
    - resident $\leftarrow$ mutant;
  - **end**
- **end**

of feasible strategies. The fixation probability $\rho_M$ of the mutant strategy can be calculated explicitly [6],

$$\rho_M = \frac{1}{1 + \sum_{i=1}^{N-1} \prod_{k=1}^{i} \frac{\lambda_k^-}{\lambda_k^+}}. \tag{3}$$

Here, the index $k$ corresponds to the current number of players with the mutant strategy (mutants). The variables $\lambda_k^-, \lambda_k^+$ are the probabilities that the number of mutants decreases or increases within a single updating step. These probabilities depend on the probability that a mutant and a resident are chosen as the learner and the resident, respectively. In addition, they depend on the respective switching probability $\varphi$, as described by Eq. (2). We specify the exact values of $\lambda_k^-, \lambda_k^+$ for each memory-setting in the next section.

Depending on the fixation probability $\rho_M$, the mutant strategy either fixes (becomes the new resident) or goes extinct. Afterwards, another random mutant strategy is introduced into the population. We iterate this elementary population updating process for a large number of mutant strategies. At each step, we record the current resident strategy and the resulting average cooperation rate.

We consider this limit of rare-mutation throughout the main text. The respective process is summarised by Algorithm 1. In Section 3, we present additional simulation results to show that our qualitative results continue to hold when the mutation rate is strictly bounded away from zero.

## 2 Analytical results

In the following, we discuss our different memory settings in more detail. We discuss six cases explicitly. In these cases, updating occurs (*i*) based on average payoffs based on all interactions (perfect memory), (*ii*) based on the last round of one interaction (limited memory), (*iii*) based on the last round of two interactions, (*iv*) based on the last two rounds of one interaction, (*v*) based on the last two rounds of two interactions, and (*v*) based on the average payoff of one interaction. In each case, we assumet there are only two strategies present in the population (a resident and a mutant strategy). We first derive how likely it is that a learner (of any type) assigns a given payoff $\pi_\text{L}$ to itself, and a payoff of $\pi_\text{RM}$ to the role model. This allows us to derive explicit expressions for $\lambda_k^-/\lambda_k^+$, and hence for the mutant's fixation probability according to Eq. (3). Based on these expressions we can characterize under which conditions cooperation is stochastically stable.

### 2.1 Perfect payoff memory

**Computing the players' expected payoffs.** The case of perfect payoff memory corresponds to the classical case considered in the previous literature. Here, individuals update their strategies based on their expected payoffs, taking into account all rounds and all possible interaction partners. When players use reactive strategies (or more generally, strategies with finite memory), these expected payoffs can be computed explicitly, based on a Markov chain approach [7]. To this end, consider two players with strategies $s_1 = (y_1, p_1, q_1)$ and $s_2 = (y_2, p_2, q_2)$, respectively. In each round $t$ of the game, player 1 may get one of the four possible payoffs $R$, $S$, $T$, or $P$, as described by the general payoff matrix (1). Let $\mathbf{v}(t) = \big(v_R(t), v_S(t), v_T(t), v_P(t)\big)$ denote the respective probability distribution of observing one of these four outcomes. This probability distribution can be computed recursively. Using the shortcut notation $\bar{z} = 1 - z$ for any $z \in [0, 1]$, we get for the initial round

$$\mathbf{v}_0 := \mathbf{v}(0) = (y_1 y_2,\ y_1 \bar{y}_2,\ \bar{y}_1 y_2,\ \bar{y}_1 \bar{y}_2). \tag{4}$$

Given $\mathbf{v}(t)$, we can compute $\mathbf{v}(t+1)$ as

$$\mathbf{v}(t+1) = \mathbf{v}(t) \cdot M, \tag{5}$$

where $M$ is the transition matrix of the process,

$$M = \begin{bmatrix} p_1 p_2 & p_1 \bar{p}_2 & \bar{p}_1 p_2 & \bar{p}_1 \bar{p}_2 \\ q_1 p_2 & q_1 \bar{p}_2 & \bar{q}_1 p_2 & \bar{q}_1 \bar{p}_2 \\ p_1 q_2 & p_1 \bar{q}_2 & \bar{p}_1 q_2 & \bar{p}_1 \bar{q}_2 \\ q_1 q_2 & q_1 \bar{q}_2 & \bar{q}_1 q_2 & \bar{q}_1 \bar{q}_2 \end{bmatrix}. \tag{6}$$

4

Based on this recursion, we can compute how often player 1 receives one of the four payoffs $R, S, T, P$ on average (across all possible realizations of games among the two players). This average distribution $\mathbf{v}$ is

$$\mathbf{v} := (1-\delta)\sum_{t=0}^{\infty}\delta^t \mathbf{v}(t) = (1-\delta)\mathbf{v_0}\sum_{t=0}^{\infty}\delta^t M^t = (1-\delta)\mathbf{v_0}(I_4-\delta M)^{-1}, \tag{7}$$

where $I_4$ is the $4\times 4$ identity matrix. Based on this general formula, the four entries of $\mathbf{v}=(v_R, v_S, v_T, v_P)$ can be computed explicitly. Using the auxiliary notation $r_i := p_i - q_i$, we obtain

$$v_R = (1-\delta)\frac{y_1 y_2}{1-\delta^2 r_1 r_2} + \delta\frac{\Big(q_1 + r_1\big((1-\delta)y_2 + \delta q_2\big)\Big)\Big(q_2 + r_2\big((1-\delta)y_1 + \delta q_1\big)\Big)}{(1-\delta r_1 r_2)(1-\delta^2 r_1 r_2)},$$

$$v_S = (1-\delta)\frac{y_1 \bar{y}_2}{1-\delta^2 r_1 r_2} + \delta\frac{\Big(q_1 + r_1\big((1-\delta)y_2 + \delta q_2\big)\Big)\Big(\bar{q}_2 - r_2\big((1-\delta)y_1 + \delta p_1\big)\Big)}{(1-\delta r_1 r_2)(1-\delta^2 r_1 r_2)},$$

$$\tag{8}$$

$$v_T = (1-\delta)\frac{\bar{y}_1 y_2}{1-\delta^2 r_1 r_2} + \delta\frac{\Big(\bar{q}_1 - r_1\big((1-\delta)y_2 + \delta p_2\big)\Big)\Big(q_2 + r_2\big((1-\delta)y_1 + \delta q_1\big)\Big)}{(1-\delta r_1 r_2)(1-\delta^2 r_1 r_2)},$$

$$v_P = (1-\delta)\frac{\bar{y}_1 \bar{y}_2}{1-\delta^2 r_1 r_2} + \delta\frac{\Big(\bar{q}_1 - r_1\big((1-\delta)y_2 + \delta p_2\big)\Big)\Big(\bar{q}_2 - r_2\big((1-\delta)y_1 + \delta p_1\big)\Big)}{(1-\delta r_1 r_2)(1-\delta^2 r_1 r_2)}.$$

Using this distribution $\mathbf{v}$, we compute the first player's expected payoff as the weighted average

$$\pi(s_1, s_2) = R\, v_R + S\, v_S + T\, v_T + P\, v_P. \tag{9}$$

The second player's payoff can be computed analogously.

**Computing the ratio $\lambda_k^-/\lambda_k^+$.** After these preparations, consider now a population with $k$ mutants and $N-k$ residents, whose strategies we denote by $s_\text{M} = (y_\text{M}, p_\text{M}, q_\text{M})$ and $s_\text{R} = (y_\text{R}, p_\text{R}, q_\text{R})$, respectively. Assuming that population members are matched randomly (or equivalently, that they interact with all other population members), the resulting expected payoffs of residents and mutants are

$$\begin{aligned}
\pi_\text{R}(k) &= \frac{N-k-1}{N-1}\cdot \pi(s_\text{R}, s_\text{R}) + \frac{k}{N-1}\cdot \pi(s_\text{R}, s_\text{M}), \\
\pi_\text{M}(k) &= \frac{N-k}{N-1}\cdot \pi(s_\text{M}, s_\text{R}) + \frac{k-1}{N-1}\cdot \pi(s_\text{M}, s_\text{M}).
\end{aligned} \tag{10}$$

The number of mutants in the population decreases in a single time step if a mutant is chosen to be the learner

and adopts the strategy of a resident. Similarly, it increases if a resident is the learner and adopts the strategy of a mutant. The respective transition probabilities are

$$\lambda_k^- = \frac{N-k}{N}\frac{k}{N-1}\varphi\big(\pi_M(k),\,\pi_R(k)\big) \quad \text{and} \quad \lambda_k^+ = \frac{k}{N}\frac{N-k}{N-1}\varphi\big(\pi_R(k),\,\pi_M(k)\big).$$

For $\varphi$ as defined by Eq. (2), the ratio of these two transition probabilities simplifies to

$$\frac{\lambda_k^-}{\lambda_k^+} = \frac{\varphi\big(\pi_M(k),\,\pi_R(k)\big)}{\varphi\big(\pi_R(k),\,\pi_M(k)\big)} = e^{-\beta\big(\pi_M(k)-\pi_R(k)\big)}. \tag{11}$$

Based on these ratios for each $k$, we can compute the mutant's fixation probability by Eq. (3).

**Stochastic stability of cooperation.** As an application of this formalism, we can compute when cooperation is stochastically stable in the perfect-information setting. To this end, suppose there is a single mutant, $k=1$. The residents adopt Generous Tit-for-Tat, $\mathtt{GTFT}=(1,1,q)$ and the mutant adopts $\mathtt{ALLD}=(0,0,0)$. When two GTFT players interact, the resulting average distribution according to Eq. (8) simplifies to

$$\mathbf{v}(\mathtt{GTFT},\mathtt{GTFT}) = (1,0,0,0).$$

On the other hand, if $\mathtt{ALLD}$ interacts with $\mathtt{GTFT}$, the respective probabilities become,

$$\mathbf{v}(\mathtt{ALLD},\mathtt{GTFT}) = \big(0,\,0,\,1-\delta+\delta q,\,\delta(1-q)\big).$$

Based on Eq. (10), we can compute the strategies' expected payoffs as

$$\pi_{\mathtt{GTFT}} = \frac{N-2}{N-1}(b-c) - \frac{1}{N-1}\big(1-\delta+\delta q\big)c \quad \text{and} \quad \pi_{\mathtt{ALLD}} = \big(1-\delta+\delta q\big)b.$$

As a consequence, we can calculate the corresponding ratio of transition probabilities according to Eq. (11),

$$\frac{\lambda_1^-}{\lambda_1^+} = e^{-\beta\big((1-\delta+\delta q)(b+\frac{c}{N-1})-\frac{N-2}{N-1}(b-c)\big)}$$

By definition, cooperation is stochastically stable if this ratio exceeds one, which is equivalent to

$$q < 1 - \frac{1}{\delta}\cdot\frac{b+(N-1)c}{(N-1)b+c}. \tag{12}$$

For such a strategy to be feasible we require $q>0$, which implies $\delta > \big(b+(N-1)c\big)/\big((N-1)b+c\big)$. In particular, in the limit of large populations $N\to\infty$, we obtain that cooperation is stochastically stable if $q<1-c/(\delta b)$. The minimum continuation probability for such a strategy to exist is $\delta > c/b$. In this way, we

6

recover the classical conditions for cooperation to be feasible under direct reciprocity [8–10].

## 2.2 Limited payoff memory

**Computing the distribution of last-round payoffs.** The case of perfect payoff memory is straightforward to handle; here, every player gets the expected payoff with certainty. In comparison, computing transition probabilities for the case of limited payoff memory is more elaborate. Here, we need to consider the different possible outcomes that both the learner and role model may have experienced in their very last interaction. A further complication arises when the learner's last interaction partner happens to be the role model. In that case, the learner's and the role model's last payoff will be correlated (e.g., if the learner got the sucker's payoff of $S$, the role model's payoff is $T$ with certainty). To treat the case of limited memory analytically, let $\mathcal{U} = \{R, S, T, P\}$ be the set of possible one-shot payoffs. Since the game ends after round $\tau$ with probability $\delta^\tau (1-\delta)$, if two players use reactive strategies $s_1 = (y_1, p_1, q_1)$ and $s_2 = (y_2, p_2, q_2)$, then, by definition, the probability that the first player receives a payoff of $u \in \mathcal{U}$ in the final round of the game is given by Eq. (8). To make the strategies explicit, we denote this probability by $v_u(s_1, s_2)$.

**Computing the ratio $\lambda_k^-/\lambda_k^+$.** After these preparations, let us again consider the corresponding population setup, with $N-k$ residents with strategy $s_\text{R} = (y_\text{R}, p_\text{R}, q_\text{R})$ and $k$ mutants with strategy $s_\text{M} = (y_\text{M}, p_\text{M}, q_\text{M})$. At each step of the evolutionary process we choose a learner and a role model. The learner compares the performance of its strategy by comparing its own last one-shot payoff with the last one-shot payoff of the role model. In the following, we assume either the learner or the role model is a resident, and that the other player is a mutant (otherwise the number of mutants does not change). There are two major cases to consider.

1. The learner and the role model have their last respective interaction with each other. This happens with probability $1/(N-1)$. In that case, there are four possible cases for their joint final payoffs, $(u_\text{R}, u_\text{M}) \in \mathcal{U}_F := \{(R, R), (S, T), (T, S), (P, P)\}$. By Eq. (8), these four outcomes follow the distribution $\mathbf{v}(s_\text{R}, s_\text{M})$.

2. The learner's last interaction was not with the role model, with probability $(N-2)/(N-1)$. In this case, there are four different subcases, depending on whether the resident's last interaction partner was a mutant or a resident, and depending on whether the mutant's last interaction partner was a mutant or a resident. As a result, the resident's last one-shot payoff is distributed according to $\mathbf{v}(s_\text{R}, s_\text{M})$ or $\mathbf{v}(s_\text{R}, s_\text{R})$; the mutant's last payoff is distributed according to $\mathbf{v}(s_\text{M}, s_\text{M})$ or $\mathbf{v}(s_\text{M}, s_\text{R})$, respectively.

Let $x(u_\text{R}, u_\text{M})$ denote the probability that the resident and the mutant received the payoff $u_\text{R}$ and $u_\text{M}$ in their

7

respective last interaction. By taking into account the above two cases, we can compute this probability as

$$x(u_R, u_M) = \tfrac{1}{N-1} \cdot v_{u_R}(s_R, s_M) \cdot 1_{(u_R, u_M) \in \mathcal{U}_F}$$

$$+ \tfrac{N-2}{N-1} \cdot \left[ \tfrac{k-1}{N-2} \tfrac{k-2}{N-3} \cdot v_{u_R}(s_R, s_M) \, v_{u_M}(s_M, s_M) + \tfrac{k-1}{N-2} \tfrac{N-k-1}{N-3} \cdot v_{u_R}(s_R, s_M) \, v_{u_M}(s_M, s_R) \right. \quad (13)$$

$$\left. + \tfrac{N-k-1}{N-2} \tfrac{k-1}{N-3} \cdot v_{u_R}(s_R, s_R) \, v_{u_M}(s_M, s_M) + \tfrac{N-k-1}{N-2} \tfrac{N-k-2}{N-3} \cdot v_{u_R}(s_R, s_R) \, v_{u_M}(s_M, s_R) \right].$$

The first line on the right side corresponds to the case that the learner and the role model happened to be matched directly for their last interaction. In that case, only those payoff pairs can occur that are feasible in a direct interaction. That is, it needs to be the case that $(u_R, u_M) \in \mathcal{U}_F$, as represented by the respective indicator function. The second and third line of (13) summarize the four possible subcases that can occur when the learner and the role model were not directly matched for their last interaction.

Overall, we now obtain the following expressions for the probability that the number of mutants increases or decreases by one,

$$\lambda_k^+ = \frac{N-k}{N} \cdot \frac{k}{N} \cdot \sum_{u_R, u_M \in \mathcal{U}} x(u_R, u_M) \cdot \varphi(u_R, u_M),$$

$$\lambda_k^- = \frac{N-k}{N} \cdot \frac{k}{N} \cdot \sum_{u_R, u_M \in \mathcal{U}} x(u_R, u_M) \cdot \varphi(u_M, u_R).$$

(14)

In this expression, the prefactor $(N-k)k/N^2$ gives the probability that the two players (learner and role model) have different strategies. The sum corresponds to the total probability that the learner adopts the role model's strategy, by summing up over all possible payoffs $u_R$ and $u_M$ that the two players may have received in their respective last rounds.

**Stochastic stability of cooperation.** To illustrate this formalism, we again use it to characterize the stability of cooperation. There is $k=1$ mutant with strategy ALLD. The remaining players use the resident strategy GTFT. When two residents interact, it follows by Eq. (8) that the outcome of the final round is distributed according to the distribution

$$\Big( v_R(\text{GTFT}, \text{GTFT}),\ v_S(\text{GTFT}, \text{GTFT}),\ v_T(\text{GTFT}, \text{GTFT}),\ v_P(\text{GTFT}, \text{GTFT}) \Big) = (1, 0, 0, 0).$$

Similarly, if the mutant interacts with a resident, the final round's outcome is distributed according to

$$\Big( v_R(\text{ALLD}, \text{GTFT}),\ v_S(\text{ALLD}, \text{GTFT}),\ v_T(\text{ALLD}, \text{GTFT}),\ v_P(\text{ALLD}, \text{GTFT}) \Big) = \big(0,\ 0,\ 1-\delta+\delta q,\ \delta(1-q)\big).$$

Based on these, we compute the probability $x(u_R, u_M)$ that the payoff of a randomly chosen GTFT player

is $u_\text{R}$ and that the payoff of the ALLD player is $u_\text{M}$, with $u_\text{R}, u_\text{M} \in \mathcal{U}$. We obtain

$$x(R,T) = \tfrac{N-2}{N-1} \cdot (1-\delta+\delta q) \qquad x(R,P) = \tfrac{N-2}{N-1} \cdot \delta(1-q)$$

$$x(S,T) = \tfrac{1}{N-1} \cdot (1-\delta+\delta q) \qquad x(P,P) = \tfrac{1}{N-1} \cdot \delta(1-q)$$

$$x(u_\text{R}, u_\text{M}) = 0 \text{ for all other payoff pairs } (u_\text{R}, u_\text{M}).$$

Based on these expressions, we calculate the ratio of transition probabilities as

$$\frac{\lambda_1^-}{\lambda_1^+} = \frac{\frac{N-2}{N-1}\left(\frac{1-\delta+\delta q}{1+e^{\beta c}} + \frac{\delta(1-q)}{1+e^{-\beta(b-c)}}\right) + \frac{1}{N-1}\left(\frac{1-\delta+\delta q}{1+e^{\beta(b+c)}} + \frac{\delta(1-q)}{2}\right)}{\frac{N-2}{N-1}\left(\frac{1-\delta+\delta q}{1+e^{-\beta c}} + \frac{\delta(1-q)}{1+e^{\beta(b-c)}}\right) + \frac{1}{N-1}\left(\frac{1-\delta+\delta q}{1+e^{-\beta(b+c)}} + \frac{\delta(1-q)}{2}\right)}.$$

Cooperation is stochastically stable if $\lambda_1^-/\lambda_1^+ > 1$. While one can solve this inequality for $q$, the resulting condition is somewhat lengthy. To obtain a more interpretable condition, we consider the limit of strong selection $\beta \to \infty$ and large populations $N \to \infty$. In that case, because $b > c > 0$, the above ratio simplifies to

$$\frac{\lambda_1^-}{\lambda_1^+} = \frac{\delta(1-q)}{1-\delta+\delta q}. \tag{15}$$

This ratio exceeds one if $q < 1 - 1/(2\delta)$. For such a strategy to be feasible, we require $q > 0$, which in turn implies $\delta > 1/2$. Moreover, in the special case that games are infinitely repeated, $\delta \to 1$, we conclude that cooperation is stochastically stable if $q < 1/2$. (For $q = 1/2$, the payoff of the ALLD player is $T > R$ for half of the time, and it is $P < R$ for the other half. The probability that the number of mutants increase by one equals the probability that the mutant goes extinct).

## 2.3 Updates based on the final round of two repeated games

**Motivation.** In the limited-memory setting, individuals take into account the outcome of one round, and of one interaction. This setting can be generalized such that an individual considers $m$ rounds and $n$ interactions. Here, we discuss the case that the update depends on the last round of $n$ interactions.

At each step of the evolutionary process, we consider the role model's and the learner's last $n$ matches. We need to define the probability that for each of the matches they are paired with a mutant, with a resident or with each other. We assume that each pair is unique, such that the learner and the role model can be matched at most once, which is a reasonable assumption in large populations. The case of $n=1$ corresponds to the previous setting of limited memory. There, we have seen that there are five possible combinations to consider. As we increase $n$, the number of possible combinations increases non-linearly; for a graphical illustration, see Supplementary Fig. 1. In the following, we study the case of $n = 2$, such that the learner takes into account the players' final payoffs of two interactions.

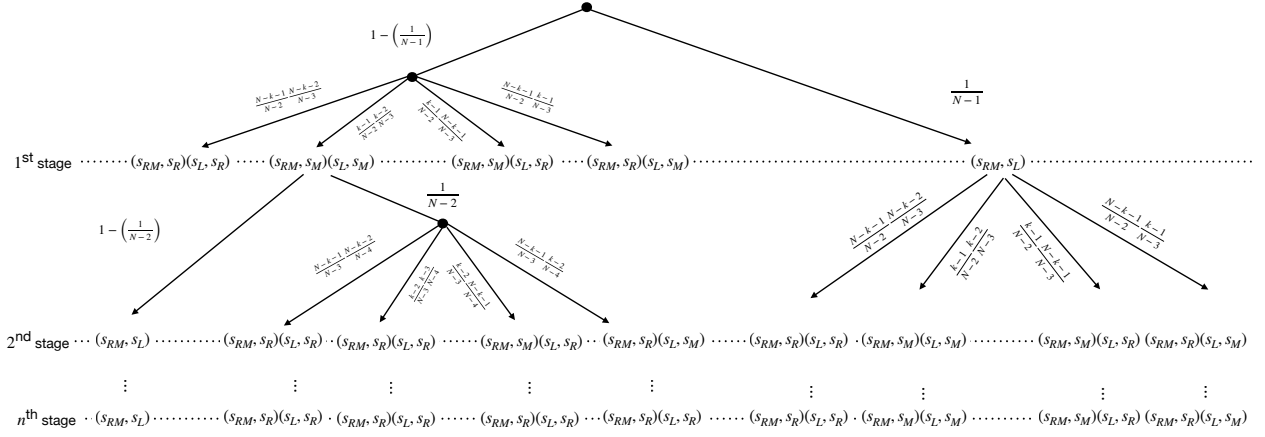

**Supplementary Figure 1: Possible pairs when the learner takes into account $n$ interactions against different co-players.** In this diagram, $(s_i, s_j)$ represents a possible pairing of two players, lines indicate possible cases, and the fractions represent the respective probabilities for each case. We consider $n$ stages in which the population members are consecutively paired with each other to play a repeated game. In the first stage, we need to consider five possible cases, as in Section 2.2. One case arises if the learner and the role model are matched directly. This happens with probability $1/(N-1)$. If they are not matched directly, both can be paired with a mutant, with a resident, or one is paired with a mutant whilst the other is paired with a resident. The second stage is similar. However, here we need to take into account that the learner and the role model can only be matched directly if they have not already interacted during the previous stage. The process continues until the $n$-th stage.

**Computing the ratio $\lambda_k^- / \lambda_k^+$.** As before, we assume that either the learner or the role model is a resident, and that the other player involved in the pairwise comparison is a mutant. For $n=2$, there are 24 possible combinations to consider. At the first stage there are five possible combinations. These are the same as in the previous setting. The learner either interacts directly with the role model or we need to take into account whether the resident interacts with a resident or a mutant, and whether the mutant interacts with a resident or a mutant. In case the learner and the role model did not directly interact with each other during the first stage, there are again five possible combinations in the second stage. Otherwise, if there already was a direct interaction, there are four combinations. Hence, there are $4 \cdot 5 + 4 = 24$ combinations in total.

We assume the resident receives the payoff $u_{R1}$ with their first interaction and $u_{R2}$ with their second. Similarly, the mutant receives payoffs $u_{M1}$ and $u_{M2}$. Let $x(u_{R1}u_{R1}, u_{M1}u_{M2})$ denote the probability that the resident and the mutant received payoffs $u_{R1}$, $u_{R2}$ and $u_{M1}$, $u_{M2}$ in the last round of their respective last two

interactions. This probability can be computed as follows,

$$
\begin{aligned}
x(u_{\text{R}1}u_{\text{R}2}, u_{\text{M}1}u_{\text{M}2}) &= \frac{1}{N-1} \cdot v_{u_{\text{R}1}}(s_{\text{R}}, s_{\text{M}}) \cdot 1_{(u_{\text{R}1}, u_{\text{M}1}) \in \mathcal{U}_F} \cdot A \\
&+ \frac{N-1}{N-2} \Bigg[ \frac{(k-1)(k-2)}{(N-2)(N-3)} v_{u_{\text{R}1}}(s_{\text{R}}, s_{\text{M}}) v_{u_{\text{M}1}}(s_{\text{M}}, s_{\text{M}}) \left( \frac{v_{u_{\text{R}2}}(s_{\text{R}}, s_{\text{M}})}{N-2} 1_{(u_{\text{R}2}, u_{\text{M}2}) \in \mathcal{U}_F} + \frac{N-3}{N-2}[B_1 + B_2 + B_3 + B_4] \right) \\
&+ \frac{(k-1)(N-k-1)}{(N-2)(N-3)} v_{u_{\text{R}1}}(s_{\text{R}}, s_{\text{M}}) v_{u_{\text{M}1}}(s_{\text{M}}, s_{\text{R}}) \left( \frac{v_{u_{\text{R}2}}(s_{\text{R}}, s_{\text{M}})}{N-2} 1_{(u_{\text{R}2}, u_{\text{M}2}) \in \mathcal{U}_F} + \frac{N-3}{N-2}[C_1 + C_2 + C_3 + C_4] \right) \quad (16)\\
&+ \frac{(k-1)(N-k-1)}{(N-2)(N-3)} v_{u_{\text{R}1}}(s_{\text{R}}, s_{\text{R}}) v_{u_{\text{M}1}}(s_{\text{M}}, s_{\text{M}}) \left( \frac{v_{u_{\text{R}2}}(s_{\text{R}}, s_{\text{M}})}{N-2} 1_{(u_{\text{R}2}, u_{\text{M}2}) \in \mathcal{U}_F} + \frac{N-3}{N-2}[D_1 + D_2 + D_3 + D_4] \right) \\
&+ \frac{(N-k-2)(N-k-1)}{(N-2)(N-3)} v_{u_{\text{R}1}}(s_{\text{R}}, s_{\text{R}}) v_{u_{\text{M}1}}(s_{\text{M}}, s_{\text{R}}) \left( \frac{v_{u_{\text{R}2}}(s_{\text{R}}, s_{\text{M}})}{N-2} 1_{(u_{\text{R}2}, u_{\text{M}2}) \in \mathcal{U}_F} + \frac{N-3}{N-2}[E_1 + E_2 + E_3 + E_4] \right) \Bigg].
\end{aligned}
$$

The first line on the right side corresponds to the case that the learner and the role model are matched during the first stage. The factor $A$ considers the different pairings that can then occur in the second stage,

$$
A = \frac{N-2}{N-1} \Bigg[ \frac{k-1}{N-2} \frac{k-2}{N-3} v_{u_{\text{M}1}}(s_{\text{R}}, s_{\text{M}}) v_{u_{\text{M}2}}(s_{\text{M}}, s_{\text{M}}) + \frac{k-1}{N-2} \frac{N-k-1}{N-3} v_{u_{\text{M}1}}(s_{\text{R}}, s_{\text{M}}) v_{u_{\text{M}2}}(s_{\text{M}}, s_{\text{R}}) \\
+ \frac{N-k-1}{N-2} \frac{k-1}{N-3} v_{u_{\text{M}1}}(s_{\text{R}}, s_{\text{R}}) v_{u_{\text{M}2}}(s_{\text{M}}, s_{\text{M}}) + \frac{N-k-1}{N-2} \frac{N-k-2}{N-3} v_{u_{\text{M}1}}(s_{\text{R}}, s_{\text{R}}) v_{u_{\text{M}2}}(s_{\text{M}}, s_{\text{R}}) \Bigg].
$$

The second line of Eq. (16) corresponds to the case that during the first stage, both the resident and the mutant interact with a mutant. In the second stage, they may interact with each other. If they do not, then again either both of them interact with a mutant, one of them interacts with a resident whereas the other interacts with a mutant, or both interact with a resident. These cases are captured by the variables

$$
B_1 = \frac{(k-2)(k-3)}{(N-3)(N-4)} v_{u_{\text{R}2}}(s_{\text{R}}, s_{\text{M}}) v_{u_{\text{M}2}}(s_{\text{M}}, s_{\text{M}}), \qquad B_2 = \frac{(k-2)(N-k-1)}{(N-3)(N-4)} v_{u_{\text{R}2}}(s_{\text{R}}, s_{\text{R}}) v_{u_{\text{M}2}}(s_{\text{M}}, s_{\text{M}}),
$$

$$
B_3 = \frac{(k-2)(N-k-1)}{(N-3)(N-4)} v_{u_{\text{R}2}}(s_{\text{R}}, s_{\text{M}}) v_{u_{\text{M}2}}(s_{\text{M}}, s_{\text{R}}), \qquad B_4 = \frac{(N-k-2)(N-k-1)}{(N-3)(N-4)} v_{u_{\text{R}2}}(s_{\text{R}}, s_{\text{R}}) v_{u_{\text{M}2}}(s_{\text{M}}, s_{\text{R}}).
$$

The third line of Eq. (16) corresponds to the case that during the first stage, the resident interacts with a mutant whereas the mutant interacts with a resident. The variables for the pairings in the second stage are

$$
C_1 = \frac{(k-3)(k-1)}{(N-3)(N-4)} v_{u_{\text{R}2}}(s_{\text{R}}, s_{\text{M}}) v_{u_{\text{M}2}}(s_{\text{M}}, s_{\text{M}}), \qquad C_2 = \frac{(k-1)(N-k-1)}{(N-3)(N-4)} v_{u_{\text{R}2}}(s_{\text{R}}, s_{\text{R}}) v_{u_{\text{M}2}}(s_{\text{M}}, s_{\text{M}}),
$$

$$
C_3 = \frac{(k-2)(N-k-2)}{(N-3)(N-4)} v_{u_{\text{R}2}}(s_{\text{R}}, s_{\text{M}}) v_{u_{\text{M}2}}(s_{\text{M}}, s_{\text{R}}), \qquad C_4 = \frac{(N-k-2)^2}{(N-3)(N-4)} v_{u_{\text{R}2}}(s_{\text{R}}, s_{\text{R}}) v_{u_{\text{M}2}}(s_{\text{M}}, s_{\text{R}}).
$$

The fourth line represents the case that during the first stage, the resident interacts with another resident whereas the mutant interacts with a mutant. The respective variables for the second-stage pairings are

$$
D_1 = \frac{(k-2)^2}{(N-3)(N-4)} v_{u_{\text{R}2}}(s_{\text{R}}, s_{\text{M}}) v_{u_{\text{M}2}}(s_{\text{M}}, s_{\text{M}}), \qquad D_2 = \frac{(k-2)(N-k-2)}{(N-3)(N-4)} v_{u_{\text{R}2}}(s_{\text{R}}, s_{\text{R}}) v_{u_{\text{M}2}}(s_{\text{M}}, s_{\text{M}}),
$$

$$
D_3 = \frac{(k-1)(N-k-1)}{(N-3)(N-4)} v_{u_{\text{R}2}}(s_{\text{R}}, s_{\text{M}}) v_{u_{\text{M}2}}(s_{\text{M}}, s_{\text{R}}), \qquad D_4 = \frac{(N-k-3)(N-k-1)}{(N-3)(N-4)} v_{u_{\text{R}2}}(s_{\text{R}}, s_{\text{R}}) v_{u_{\text{M}2}}(s_{\text{M}}, s_{\text{R}}).
$$

Finally, the last line of Eq. (16) corresponds to the case that during the first stage, both the resident and the mutant interact with a resident. The variables for the second stage are

$$E_1 = \frac{(k-2)(k-1)}{(N-3)(N-4)}\, v_{u_{R2}}(s_R, s_M)\, v_{u_{M2}}(s_M, s_M), \qquad E_2 = \frac{(k-1)(N-k-2)}{(N-3)(N-4)}\, v_{u_{R2}}(s_R, s_R)\, v_{u_{M2}}(s_M, s_M),$$

$$E_3 = \frac{(k-1)(N-k-2)}{(N-3)(N-4)}\, v_{u_{R2}}(s_R, s_M)\, v_{u_{M2}}(s_M, s_R), \qquad E_4 = \frac{(N-k-3)(N-k-2)}{(N-3)(N-4)}\, v_{u_{R2}}(s_R, s_R)\, v_{u_{M2}}(s_M, s_R).$$

As a result, the transition probabilities for the number of mutants is now given by

$$\begin{aligned}
\lambda_k^+ &= \frac{N-k}{N} \cdot \frac{k}{N} \sum_{u_{R1},u_{R2},u_{M1},u_{M2}} x(u_{R1}u_{R2}, u_{M1}u_{M2}) \cdot \varphi\left(\frac{u_{R1}+u_{R2}}{2}, \frac{u_{M1}+u_{M2}}{2}\right), \\
\lambda_k^- &= \frac{N-k}{N} \cdot \frac{k}{N} \sum_{u_{R1},u_{R2},u_{M1},u_{M2}} x(u_{R1}u_{R2}, u_{M1}u_{M2}) \cdot \varphi\left(\frac{u_{M1}+u_{M2}}{2}, \frac{u_{R1}+u_{R2}}{2}\right).
\end{aligned} \qquad (17)$$

Note that we assume players to use their average payoff across their two interactions for comparisons.

**Stochastic stability of cooperation.** As before, we can use this formalism to compute when cooperation is stochastically stable. For a single `ALLD` mutant in a `GTFT` resident population, we obtain the following non-zero probabilities $x(u_{R1}u_{R2}, u_{M1}, u_{M2})$,

$$\begin{aligned}
x(RR,TT) &= \tfrac{N-3}{N-1}(\delta q - \delta + 1)^2, & x(RR,TP) &= \tfrac{N-3}{N-1}\delta(1-q)(\delta q - \delta + 1), \\
x(RR,PT) &= \tfrac{N-3}{N-1}\delta(1-q)(\delta q - \delta + 1), & x(RR,PP) &= \tfrac{N-3}{N-1}\delta^2(1-q)^2, \\
x(RS,TT) &= \tfrac{1}{N-1}(\delta q - \delta + 1)^2, & x(RS,PT) &= \tfrac{1}{N-1}\delta(1-q)(\delta q - \delta + 1), \\
x(RP,TP) &= \tfrac{1}{N-1}\delta(1-q)(\delta q - \delta + 1), & x(RP,PP) &= \tfrac{1}{N-1}\delta^2(1-q)^2, \\
x(SR,TT) &= \tfrac{1}{N-1}(\delta q - \delta + 1)^2, & x(SR,TP) &= \tfrac{1}{N-1}\delta(1-q)(\delta q - \delta + 1), \\
x(PR,PT) &= \tfrac{1}{N-1}\delta(1-q)(\delta q - \delta + 1), & x(PR,PP) &= \tfrac{1}{N-1}\delta^2(1-q)^2.
\end{aligned}$$

The resulting ratio of transition probabilities is

$$\frac{\lambda_1^-}{\lambda_1^+} = \frac{\frac{N-3}{N-1}\left(\frac{\delta^2(1-q)^2}{1+e^{-\beta(b-c)}} + \frac{2\delta(1-q)(\delta q-\delta+1)}{1+e^{-\beta(b/2-c)}} + \frac{(\delta q-\delta+1)^2}{1+e^{\beta c}}\right) + \frac{2}{N-1}\left(\frac{\delta^2(1-q)^2}{1+e^{-\beta(b-c)/2}} + \frac{\delta(1-q)(\delta q-\delta+1)}{1+e^{-\beta c/2}} + \frac{\delta(1-q)(\delta q-\delta+1)}{1+e^{\beta c}} + \frac{(\delta q-\delta+1)^2}{1+e^{\beta(b/2+c)}}\right)}{\frac{N-3}{N-1}\left(\frac{\delta^2(1-q)^2}{1+e^{\beta(b-c)}} + \frac{2\delta(1-q)(\delta q-\delta+1)}{1+e^{\beta(b/2-c)}} + \frac{(\delta q-\delta+1)^2}{1+e^{-\beta c}}\right) + \frac{2}{N-1}\left(\frac{\delta^2(1-q)^2}{1+e^{\beta(b-c)/2}} + \frac{\delta(1-q)(\delta q-\delta+1)}{1+e^{-\beta c/2}} + \frac{\delta(1-q)(\delta q-\delta+1)}{1+e^{-\beta c}} + \frac{(\delta q-\delta+1)^2}{1+e^{-\beta(b/2+c)}}\right)}. \qquad (18)$$

Again this expression simplifies considerably in the limit of strong selection, $\beta \to \infty$, and of large popula-

tions, $N \to \infty$. In that case, we obtain the following three cases

$$\frac{\lambda_1^-}{\lambda_1^+} = \begin{cases} \frac{1-(1-\delta(1-q))^2}{(1-\delta(1-q))^2} & \text{if } \frac{b}{2} > c, \\ \frac{\delta(1-q)}{1-\delta(1-q)} & \text{if } \frac{b}{2} = c, \\ \frac{\delta^2(1-q)^2}{1-\delta^2(1-q)^2} & \text{if } \frac{b}{2} < c. \end{cases} \tag{19}$$

Unlike in the case of limited payoff memory, here the exact magnitude of the benefit and the cost of cooperation do have some effect on the transition probabilities. By requiring that the above ratio exceeds one, we obtain the following conditions for stochastic stability of cooperation,

$$\begin{cases} q < \frac{\delta - 1 + \sqrt{2}/2}{\delta} & \text{if } \frac{b}{2} > c, \\ q < 1 - \frac{1}{2\delta} & \text{if } \frac{b}{2} = c, \\ q < \frac{\delta - \sqrt{2}/2}{\delta} & \text{if } \frac{b}{2} < c. \end{cases} \tag{20}$$

Only when $b/2 = c$ we obtain the same expression as in the limited-memory case, Eq. (15). The requirement $q \geq 0$ implies that the minimum continuation probability is $\delta > 1 - \sqrt{2}/2 \approx 0.293$ in the first case, $\delta > \frac{1}{2}$ in the second, and $\delta > \sqrt{2}/2 \approx 0.707$ in the third case. Conversely, if the game is infinitely repeated, $\delta \to 1$, stochastic stability of cooperation requires $q < \sqrt{2}/2 \approx 0.707$ in the first case, $q < 1/2$ in the second, and $q < 1 - \sqrt{2}/2 \approx 0.293$ in the third case.

## 2.4 Updates based on the last two rounds of one repeated game

**General setup.** As our next setting, we assume players update their strategies based on the last two rounds of their last game. Here, additional complications may arise because some games may only last for one round if $\delta < 1$. In the following, we abstract from these complications. We shall assume that learners only update their strategies when both the learner's and the role model's last game had at least two rounds. If either the learner or the role model only engaged in a one-round game in their previous interaction, we assume the learner decides to keep its strategy. As a result, in the following, we accordingly re-interpret some of our quantities. For example, $\lambda_k^+$ is no longer the (unconditional) probability that the number of mutants increases by one after a learner and a role model have been chosen. Instead, we interpret it as the (conditional) probability that the number of mutants increases, given that the learner and the role model happened to interact at least for two rounds during their respective last interaction. This new interpretation of transition probabilities affects the time scale of fixation. After all, we ignore all time steps in which no update occurred because a player's last game only lasted for a single round. However, our re-interpretation does not affect the ratio $\lambda_k^-/\lambda_k^+$, which is the key quantity we consider.

**Computing the distribution of outcomes in the penultimate and in the final round of a game.** To compute the ratio of transition probabilities, again we need to compute how likely it is that the resident and the mutant received any possible combination of two payoffs during the last two rounds. The probability that the game lasts for at least two rounds is $\delta$, since that is the probability of continuing from round zero (the initial round) to round one. The distribution of actions in in the penultimate round, conditioned on there being at least two rounds in the game, is then

$$\frac{1}{\delta} \sum_{t=1}^{\infty} \delta^t (1-\delta) \mathbf{v_0} M^{t-1} = (1-\delta) \mathbf{v_0} \sum_{t=0}^{\infty} \delta^t M^t, \tag{21}$$

which is identical to the expression for $\mathbf{v}$ in Eq. (7). Thus, the probability that the first player receives the payoff $u \in \mathcal{U}$ in the penultimate round of the game, conditional on the event that the game lasts at least two rounds, is again identical to the player's average probability $v_u$ to receive payoff $u$ across all rounds of the game, as specified by Eq. (8).

We can now derive a probability distribution for the first player's payoff in the last two rounds of a repeated game. To this end, consider the $4 \times 4$ transition matrix $M(s_1, s_2) = (m_{u,u'})$ according to Eq. (6). Instead of the usual indexing of the four rows by numbers, $i \in \{1, 2, 3, 4\}$, here we label the four rows of this matrix by the first player's payoffs in the previous round, $u \in \{R, S, T, P\}$. Similarly, we label the four columns of this matrix by the resulting payoffs to player 1 in the next round. For example, $m_{ST}$ corresponds to the second row and third column of matrix $M$. So by Eq. (6), $m_{ST} = \bar{q}_1 p_2 = (1-q_1) p_2$. Using this notation, we can describe the probability $w_{uu'}(s_1, s_2)$ that the first player receives the payoffs $u$ and $u'$ in the last two rounds. By combining the probabilities that player 1 obtains a payoff of $u$ in the penultimate round (probability $v_u$) and then a payoff of $u'$ in the last round (conditional probability $m_{u,u'}$), we obtain $w_{uu'}(s_1, s_2) = v_u \cdot m_{u,u'}$.

**Computing the ratio $\lambda_k^-/\lambda_k^+$.** To compute the ratio of transition probabilities, let us again take a population perspective. We consider $k$ mutants with strategy $s_\text{M} = (y_\text{M}, p_\text{M}, q_\text{M})$ and $N-k$ residents with strategy $s_\text{R} = (y_\text{R}, p_\text{R}, q_\text{M})$. Without loss of generality, we assume that either the learner or the role model is a resident, and that the respective other player is a mutant. Let $x(u_\text{R} u'_\text{R}, u_\text{M} u'_\text{M})$ be the probability that the two players received payoffs $u$ and $u'$ in the last two rounds of their last repeated game. We can compute this

probability based on the same logic as in the limited-memory setting (Section 2.2). This yields

$$x(u_\text{R}u'_\text{R}, u_\text{M}u'_\text{M}) = \tfrac{1}{N-1} \cdot w_{u_\text{R}u'_\text{R}}(s_\text{R}, s_\text{M}) \cdot 1_{(u_\text{R}, u_\text{M}) \in \mathcal{U}_F} \cdot 1_{(u'_\text{R}, u'_\text{M}) \in \mathcal{U}_F}$$

$$+ \tfrac{N-2}{N-1} \cdot \left[ \tfrac{k-1}{N-2} \tfrac{k-2}{N-3} \cdot w_{u_\text{R}u'_\text{R}}(s_\text{R}, s_\text{M}) w_{u_\text{M}u'_\text{M}}(s_\text{M}, s_\text{M}) + \tfrac{k-1}{N-2} \tfrac{N-k-1}{N-3} \cdot w_{u_\text{R}u'_\text{R}}(s_\text{R}, s_\text{M}) w_{u_\text{M}u'_\text{M}}(s_\text{M}, s_\text{R}) \right.$$

$$\left. + \tfrac{N-k-1}{N-2} \tfrac{k-1}{N-3} \cdot w_{u_\text{R}u'_\text{R}}(s_\text{R}, s_\text{R}) w_{u_\text{M}u'_\text{M}}(s_\text{M}, s_\text{M}) + \tfrac{N-k-1}{N-2} \tfrac{N-k-2}{N-3} \cdot w_{u_\text{R}u'_\text{R}}(s_\text{R}, s_\text{R}) w_{u_\text{M}u'_\text{M}}(s_\text{M}, s_\text{R}) \right].$$

Overall, we obtain the following formula for the transition probabilities

$$\begin{aligned} \lambda_k^+ &= \frac{N-k}{N} \cdot \frac{k}{N} \sum_{u_\text{R}, u'_\text{R}, u_\text{M}, u'_\text{M}} x(u_\text{R}u'_\text{R}, u_\text{M}u'_\text{M}) \cdot \varphi\left(\frac{u_\text{R}+u'_\text{R}}{2}, \frac{u_\text{M}+u'_\text{M}}{2}\right), \\ \lambda_k^- &= \frac{N-k}{N} \cdot \frac{k}{N} \sum_{u_\text{R}, u'_\text{R}, u_\text{M}, u'_\text{M}} x(u_\text{R}u'_\text{R}, u_\text{M}u'_\text{M}) \cdot \varphi\left(\frac{u_\text{M}+u'_\text{M}}{2}, \frac{u_\text{R}+u'_\text{R}}{2}\right). \end{aligned} \quad (22)$$

**Stochastic stability of cooperation.** We once again calculate how easily a single `ALLD` mutant can invade a resident population of `GTFT` players. When two residents interact, we obtain the following probability distribution for the outcome of the two last rounds,

$$w_{RR}(\text{GTFT}, \text{GTFT}) = 1,$$
$$w_{uu'}(\text{GTFT}, \text{GTFT}) = 0 \quad \text{for all other } u, u' \in \mathcal{U}.$$

For interactions between the mutant and a resident, we obtain

$$w_{TT}(\text{ALLD}, \text{GTFT}) = q(\delta q - \delta + 1), \qquad w_{TP}(\text{ALLD}, \text{GTFT}) = (1-q)(\delta q - \delta + 1),$$
$$w_{PT}(\text{ALLD}, \text{GTFT}) = \delta q(1-q), \qquad w_{PP}(\text{ALLD}, \text{GTFT}) = \delta(1-q)^2,$$
$$w_{uu'}(\text{ALLD}, \text{GTFT}) = 0 \text{ for all other } u, u' \in \mathcal{U}.$$

As a consequence, we obtain the following probabilities $x(u_\text{R}u'_\text{R}, u_\text{M}u'_\text{M})$ that the payoffs of a randomly chosen `GTFT` resident are $u_\text{R}$ and $u'_\text{R}$, whereas the payoffs of the mutant are $u_\text{M}$ and $u'_\text{M}$:

$$\begin{aligned} x(RR, TT) &= \tfrac{N-2}{N-1} q(\delta q - \delta + 1), & x(RR, TP) &= \tfrac{N-2}{N-1}(1-q)(\delta q - \delta + 1), \\ x(RR, PT) &= \tfrac{N-2}{N-1} \delta q(1-q), & x(RR, PP) &= \tfrac{N-2}{N-1} \delta(1-q)^2, \\ x(SS, TT) &= \tfrac{1}{N-1} q(\delta q - \delta + 1), & x(SP, TP) &= \tfrac{1}{N-1}(1-q)(\delta q - \delta + 1), \\ x(PS, PT) &= \tfrac{1}{N-1} \delta q(1-q), & x(PP, PP) &= \tfrac{1}{N-1} \delta(1-q)^2, \end{aligned}$$

$$x(u_\text{R}u'_\text{R}, u_\text{M}u'_\text{M}) = 0 \quad \text{for all other payoff combinations.}$$

Based on these probabilities, we compute the ratio

$$\frac{\lambda_1^-}{\lambda_1^+} = \frac{\frac{N-2}{N-1}\left(\frac{\delta(1-q)^2}{1+e^{-\beta(b-c)}}+\frac{q(\delta q-\delta+1)}{1+e^{\beta c}}+\frac{(1-q)(2\delta q-\delta+1)}{1+e^{-\beta(b/2-c)}}\right)+\frac{1}{N-1}\left(\frac{\delta(1-q)^2}{2}+\frac{q(\delta q-\delta+1)}{1+e^{\beta(b+c)}}+\frac{(1-q)(2\delta q-\delta+1)}{1+e^{\beta(b+c)/2}}\right)}{\frac{N-2}{N-1}\left(\frac{\delta(1-q)^2}{1+e^{\beta(b-c)}}+\frac{q(\delta q-\delta+1)}{1+e^{-\beta c}}+\frac{(1-q)(2\delta q-\delta+1)}{1+e^{\beta(b/2-c)}}\right)+\frac{1}{N-1}\left(\frac{\delta(1-q)^2}{2}+\frac{q(\delta q-\delta+1)}{1+e^{-\beta(b+c)}}+\frac{(1-q)(2\delta q-\delta+1)}{1+e^{-\beta(b+c)/2}}\right)}. \quad (23)$$

In the limit of strong selection $\beta \to \infty$ and large populations $N \to \infty$, this ratio simplifies to

$$\frac{\lambda_1^-}{\lambda_1^+} = \begin{cases} \frac{(1-q)(1+\delta q)}{q(\delta q-\delta+1)} & \text{if } \frac{b}{2} > c, \\ \frac{(1+\delta)(1-q)}{\delta q-\delta+q+1} & \text{if } \frac{b}{2} = c, \\ \frac{\delta(1-q)^2}{1-\delta(1-q)^2} & \text{if } \frac{b}{2} < c. \end{cases} \quad (24)$$

For stochastic stability of cooperation we require this ratio to be larger than one. This yields the conditions

$$\begin{cases} q < \frac{1}{1-\delta+\sqrt{1+\delta^2}} & \text{if } \frac{b}{2} > c, \\ q < \frac{\delta}{1+\delta} & \text{if } \frac{b}{2} = c, \\ q < 1 - \frac{1}{\sqrt{2\delta}} & \text{if } \frac{b}{2} < c. \end{cases} \quad (25)$$

Interestingly, these conditions differ from the respective conditions in the previous section, Eq. (20). Moreover, the requirement $q \geq 0$ imposes no constraints in the first two cases; it imposes the constraint $\delta > 1/2$ in the last case. However, in the limit of long games, $\delta \to 1$ we recover the same conditions as in the previous section. Stochastic stability of cooperation requires $q < \sqrt{2}/2 \approx 0.707$ in the first case, $q < 1/2$ in the second, and $q < 1-\sqrt{2}/2 \approx 0.293$ in the third case.

## 2.5 Updates based on the last two rounds of two repeated games

For our next extension, we consider a combination of the previous two settings, discussed in Sections 2.3 and 2.4. Now, individuals take into account two interactions with different co-players, and from each interaction, they take into account the last two rounds. The probability that the number of mutants increases or decreases

by one is now given by

$$\lambda_k^+ = \frac{N-k}{N} \cdot \frac{k}{N} \sum_{\substack{u_{R1}, u'_{R1}, u_{R2}, u'_{R2} \\ u_{M1}, u'_{M1}, u_{M2}, u'_{M2}}} x(u_{R1} u'_{R1} u_{R2} u'_{R2}, u_{M1} u'_{M1} u_{M2} u'_{M2}) \cdot \varphi\left(\frac{u_{R1}+u'_{R1}+u_{R2}+u'_{R2}}{4}, \frac{u_{M1}+u'_{M1}+u_{M2}+u'_{M2}}{4}\right),$$

$$\lambda_k^- = \frac{N-k}{N} \cdot \frac{k}{N} \sum_{\substack{u_{R1}, u'_{R1}, u_{R2}, u'_{R2} \\ u_{M1}, u'_{M1}, u_{M2}, u'_{M2}}} x(u_{R1} u'_{R1} u_{R2} u'_{R2}, u_{M1} u'_{M1} u_{M2} u'_{M2}) \cdot \varphi\left(\frac{u_{M1}+u'_{M1}+u_{M2}+u'_{M2}}{4}, \frac{u_{R1}+u'_{R1}+u_{R2}+u'_{R2}}{4}\right).$$

(26)

Due to the large number of cases to consider in this setting, we do not derive conditions for the stochastic stability of cooperation.

### 2.6 Updates based on the average payoff of one repeated game

**Motivation.** As our last setting, we consider the case that players update their strategies based on an (entire) repeated game against a single co-player. In terms of cognitive complexity, this setting is in between the case of perfect memory and the case considered in Section 2.4, where players only take into account two rounds of one repeated game.

**Computing the ratio $\lambda_k^-/\lambda_k^+$.** As in the previous sections, we consider $k$ mutants with strategy $s_M = (y_M, p_M, q_M)$ and $N-k$ residents with strategies $s_R = (y_R, p_R, q_R)$. Moreover, without loss of generality, we only consider those cases in which either the learner or the role model is a resident, and the other player is a mutant. As in previous cases, two major cases must be considered, (*i*) the learner and the role model interact directly, and (*ii*) the learner's last repeated game was not with the role model. In each case, the learner's decision is now based on the players' expected payoffs, as given by Eq. (9), with their last interaction partner. Based on these observations, we can derive the probability that the number of mutants either increases or decreases by one,

$$\lambda^+ = \tfrac{1}{N-1} \cdot \varphi\left(\pi(s_R, s_M), \pi(s_M, s_R)\right)$$
$$+ \tfrac{N-2}{N-1} \cdot \left[ \tfrac{k-1}{N-2} \tfrac{k-2}{N-3} \cdot \varphi(\pi(s_R, s_M), \pi(s_M, s_M)) + \tfrac{k-1}{N-2} \tfrac{N-k-1}{N-3} \cdot \varphi(\pi(s_R, s_M), \pi(s_M, s_R)) \right.$$
$$\left. + \tfrac{N-k-1}{N-2} \tfrac{k-1}{N-3} \cdot \varphi(\pi(s_R, s_R), \pi(s_M, s_M)) + \tfrac{N-k-1}{N-2} \tfrac{N-k-2}{N-3} \cdot \varphi(\pi(s_R, s_R), \pi(s_M, s_R)) \right],$$

$$\lambda^- = \tfrac{1}{N-1} \cdot \varphi\left(\pi(s_M, s_R), \pi(s_R, s_M)\right)$$
$$+ \tfrac{N-2}{N-1} \cdot \left[ \tfrac{k-1}{N-2} \tfrac{k-2}{N-3} \cdot \varphi(\pi(s_M, s_M), \pi(s_R, s_M)) + \tfrac{k-1}{N-2} \tfrac{N-k-1}{N-3} \cdot \varphi(\pi(s_M, s_R), \pi(s_R, s_M)) \right.$$
$$\left. + \tfrac{N-k-1}{N-2} \tfrac{k-1}{N-3} \cdot \varphi(\pi(s_M, s_M), \pi(s_R, s_R)) + \tfrac{N-k-1}{N-2} \tfrac{N-k-2}{N-3} \cdot \varphi(\pi(s_M, s_R), \pi(s_R, s_R),) \right].$$

**Stochastic stability of cooperation.** We once again calculate how easily a single `ALLD` mutant can invade into a resident population of `GTFT` players. Depending on whether a resident interacts with another resident or with a mutant, the resident's average payoff based on Eq. (9) is

$$\pi(s_\text{R}, s_\text{R}) = b - c \qquad \text{or} \qquad \pi(s_\text{R}, s_\text{M}) = -c\big(1 - \delta(1-q)\big).$$

The mutant can only interact with a resident, yielding the payoff

$$\pi(s_\text{M}, s_\text{R}) = b\big(1 + \delta\,(1-q)\big).$$

Based on these average payoffs, we compute the ratio

$$\frac{\lambda_1^-}{\lambda_1^+} = \frac{\frac{1}{N-1}\frac{1}{\left(1+e^{\beta(b(\delta q-\delta)+c)}\right)} + \frac{N-2}{N-1}\frac{1}{\left(1+e^{-\beta(-b(\delta q-\delta+1)+b-c)}\right)}}{\frac{1}{N-1}\frac{1}{\left(1+e^{-\beta(b(\delta q-\delta)+c)}\right)} + \frac{N-2}{N-1}\frac{1}{\left(1+e^{-\beta(b(\delta q-\delta+1)-b+c)}\right)}}. \tag{27}$$

In the limit of large populations $N \to \infty$, this ratio simplifies to

$$\frac{\lambda_1^-}{\lambda_1^+} = \frac{1 + e^{-\beta(b(\delta q - \delta) + c)}}{1 + e^{\beta(b(\delta q - \delta) + c)}}. \tag{28}$$

For stochastic stability of cooperation we require this ratio to be larger than one, which is equivalent to

$$q < 1 - \frac{c}{\delta b} \tag{29}$$

Interestingly, this condition is the same as condition (12) for the case of perfect memory when $N \to \infty$.

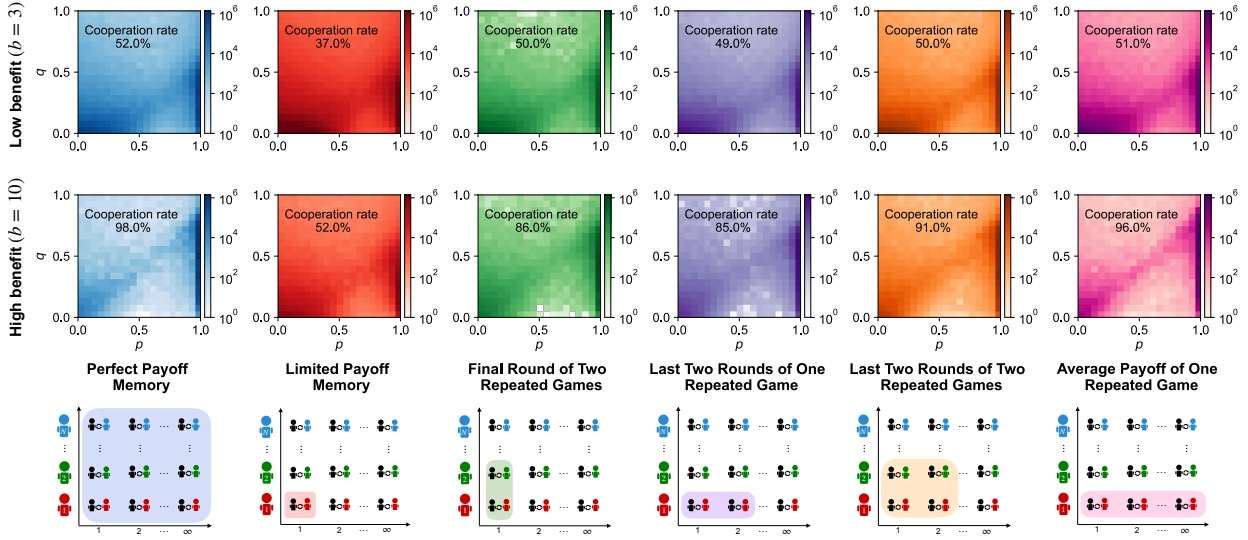

**Supplementary Figure 2: Evolutionary dynamics of our six different settings.** From left to right, we present results on the following cases: individuals update their strategies based on (*i*) the expected payoffs (perfect memory); (*ii*) the last-round payoff of one interaction (limited memory); (*iii*) the last round payoff of two interactions; (*iv*) the payoffs of the last two rounds of one interaction; (*v*) the payoff of the last two rounds of two interactions and (*vi*) the expected payoff of one interaction. Simulations are run either for a comparably high benefit of cooperation ($b/c = 10$), or for a low benefit ($b/c = 3$). We run each simulation for $T = 10^7$ time steps. For each time step we record the current resident population $(y, p, q)$. The graphs show how often the resident population chooses each combination $(p, q)$; because $\delta \approx 1$, we do not show the distribution of the player's first-round cooperation probability $y$.

## 3 Further simulation results

### 3.1 Further simulation results for reactive strategies

**Strategies favored by selection.** To gain some insights into the strategies evolving in each of the settings described in the previous section, we simulate the evolutionary process according to Algorithm 1. The results of these simulations are shown in Supplementary Fig. 2. Qualitatively, all settings lead to similar dynamics as in the case of perfect and limited memory. In particular, populations are clustered around ALLD, and around a strip of conditionally cooperative strategies, GTFT=$(1, 1, q)$. In all settings, we observe that along this strip, strategies are most abundant when they satisfy the condition for stochastic stability. For perfect memory, this means the resident population predominantly adopts a strategy with $q \leq 1 - \frac{c}{b} = 0.9$ for high benefit and $q \leq 1 - \frac{c}{b} = 0.67$ for low benefit. For limited memory, the generosity of a conditional cooperator satisfies $q < \frac{1}{2}$ for both benefit values. In the next two cases, when players either remember the last round of two interactions or the last two rounds of one interaction, stochastic stability requires $q < \frac{\sqrt{2}}{2} \approx 0.701$. In the last case, we get the same condition for the generosity of a conditional cooperator as in the perfect memory case: $q < 1 - \frac{c}{b} = 0.9$ for high benefit and $q < 1 - \frac{c}{b} = 0.67$ for low benefit. In all cases, the simulation results are well aligned with these theoretical predictions.

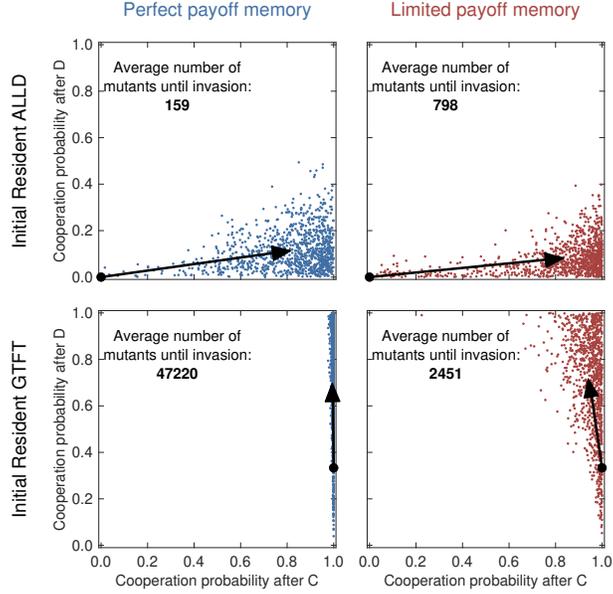

**Supplementary Figure 3: Invasion analysis for perfect and limited payoff memory.** We consider two residents, ALLD= $(0,0,0)$ and GTFT= $(1,1,1/3)$ for both perfect and limited payoff memory. In each case, we simulate the pairwise comparison process $10^3$ times and record the first mutant strategy that reaches fixation (colored dots). Black arrows indicate the arithmetic mean of the successful mutant strategies. In addition, we record how many mutant strategies it takes on average to successfully invade a given resident. We observe that under perfect memory, defectors are more prone to being invaded, whereas conditional cooperators are more resilient to invasion. Parameters are the same as in Fig. 1 of the main text, with $b=10$, $c=1$, $N=100$, $\beta=1$, $\delta=0.999$.

**Invasion dynamics.** To gain some further into the dynamics over time, and into the differences between perfect and limited payoff memory, we perform an invasion analysis, see Supplementary Fig. 3. Here, we consider two possible resident strategies. Either populations use ALLD= $(0,0,0)$ (top row) or they start out with GTFT= $(1,1,1/3)$ (bottom row). In each case, we simulate the process until a mutant reaches fixation, for both perfect memory (left) and limited memory (right). We iterate this simulation $10^3$ times. Colored dots indicate the position of successful mutants, and arrows indicate the average trajectory of evolution. Moreover, we also record how long it takes on average until the first mutant strategy takes over. This plot shows that ALLD populations are typically invaded by mutants similar to Tit-for-Tat, with $p>0$ and $q\approx 0$. In contrast, GTFT populations are typically invaded by other GTFT mutants, but with higher levels of generosity. A further analysis of the fixation time can shed some light on why perfect memory leads to higher average cooperation rates: Compared to limited memory, it takes fewer mutants to successfully invade into ALLD, and more mutants to invade into GTFT.

**Impact of the mutation rate.** All of our previous simulations are run under the assumption that mutations are rare, $\mu \to 0$. This allows us to use the formula (3) for a mutant's fixation probability. While this limit allows for more efficient computation procedures, it rules out the possibility that two or three strategies

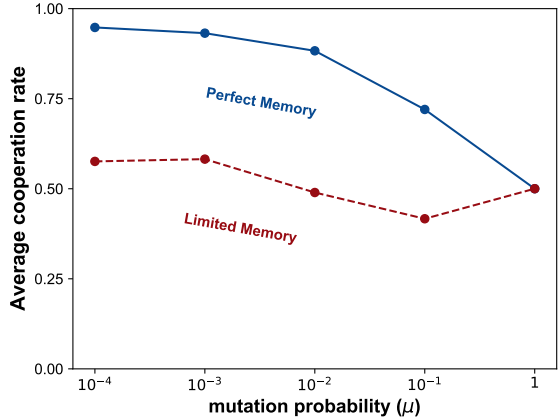

**Supplementary Figure 4: Evolutionary dynamics results for perfect and limited memory for different mutation values.** We perform five independent simulations. Simulations are run for $T = 4 \times 10^7$ time steps for each parameter. In each time step, we introduce a new mutant with a probability $\mu$, and we then select two random players to serve as the role model and the learner. The learner adopts the strategy of the role model with a probability $\varphi(\pi_L, \pi_{RM})$ where the payoffs depend on the relevant setting. We plot the average cooperation rate within the resident population for each value of $\mu$. The results show that our previous results for rare mutations hold more generally: While limited memory allows the evolution of moderate cooperation levels, perfect memory is always more conducive to the evolution of cooperation. Parameters are the same as before, $N = 100, c = 1, b = 10, \beta = 1, \delta = 0.999$.

might stably coexist in a population [11]. To explore the effect of positive mutation rates, we perform five independent runs of the pairwise process described in Section 1 for different values of $\mu$. At each time step, we record the average cooperation rate of the resident population. The results for different mutation rates are shown in Supplementary Fig. 4. For almost all considered mutation rates, the cooperation rate is higher when players have perfect memory (compared to limited memory). Only as the mutation rate approaches one, and the processes become fully random, does the cooperation rate approach $1/2$ in both simulations.

## 3.2 Simulations for memory-one strategies

So far, we have assumed that individuals adopt reactive strategies. To demonstrate that our results hold for more complex strategy sets, in the following we present results when players can choose among all memory-one strategies [7]. Players with these strategies consider the full outcome of the previous round to decide on their next action. There are four possible outcomes in each round, $(C, C), (C, D), (D, C), (D, D)$. A memory-one strategy $s$ can thus be written as a five-dimensional vector $s = (y, p_1, p_2, p_3, p_4)$. As with reactive strategies, the entry $y$ is the probability that the strategy opens with a cooperation. The other entries $p_1, p_2, p_3, p_4$ are the probabilities that the strategy cooperates in all subsequent rounds, depending on the outcome of the previous round. Every reactive strategy $(y, p, q)$ can be written as a memory-one strategy, by using the memory-one representation $(y, p, q, p, q)$. Conversely, there are memory-one strategies that cannot be represented by a reactive strategy, such as the strategy WSLS= $(1, 1, 0, 0, 1)$. Therefore, the set of

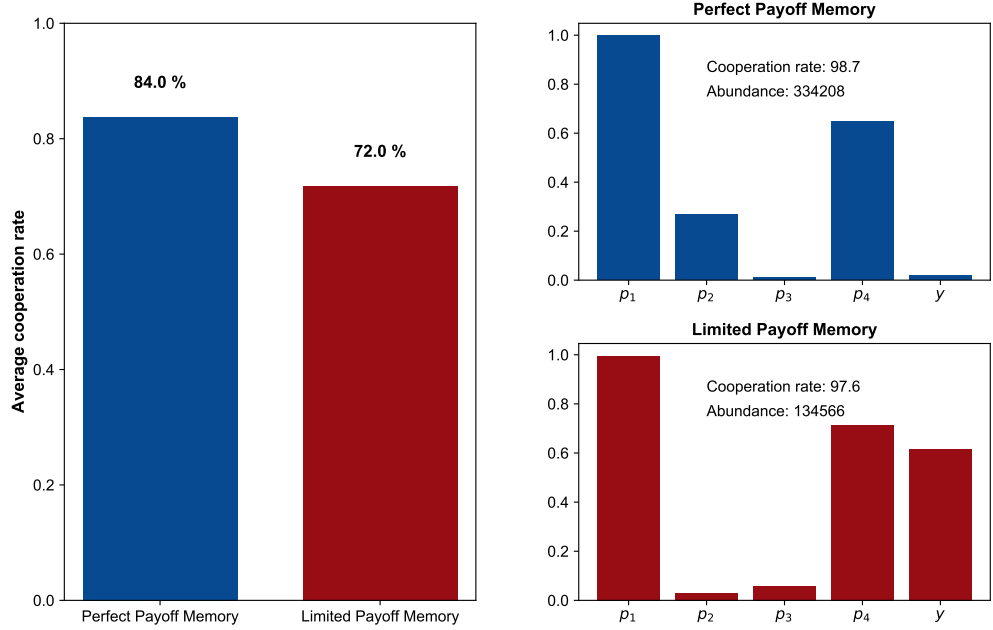

**Supplementary Figure 5: Evolution of memory-one strategies when the benefit of cooperation is comparably small.** We perform two independent simulations. In one simulation, individuals use expected payoffs. In the other, they update their strategy based on the payoff they received in the last round of their last interaction. We run each simulation for $T = 10^8$ time steps. For each time step, we record the current resident population, which is now of the form $(y, p_1, p_2, p_3, p_4)$. In the left panel, we report the cooperation rates for each simulation. For memory-one strategies, like for reactive strategies, the use of expected payoffs results in more cooperation. The right panel reports the most abundant strategy of each simulation. Abundance is the number of mutants a strategy can repel before being invaded. The most abundant strategies have some similarities, namely, $p_1 \approx 1$, $p_3 \approx 0$ and $p_4 > \frac{1}{2}$. Parameters: $N = 100, c = 1, b = 3, \beta = 1, \delta = 0.999$.

memory-one strategies is a strict superset of the reactive strategies.

To explore the dynamics among memory-one strategies, we perform four separate simulations, for both perfect and limited memory, and for two different benefit values. Supplementary Fig. 5 shows the results for a comparably low benefit of cooperation; Supplementary Fig. 6 shows the corresponding results for a large benefit. These simulations suggest that even when individuals are allowed to use memory-one strategies, there is more cooperation when players update their strategies based on perfect payoff memory.

To further investigate the influence of different parameters, we have systematically varied the benefit $b$ and the selection strength $\beta$, as shown in Supplementary Fig. 7. Similar to the case of reactive strategies, we observe that cooperation rates in the setting with limited memory remain stable after $b \geq 4$, see Supplementary Fig. 7(a). Moreover, also in this scenario, we observe that perfect memory consistently results in more cooperation. Regarding the effect of different selection strengths, as depicted in Supplementary Fig. 7(b), the results once again suggest that both perfect and limited payoff memory yield similar cooperation rates for weak selection ($\beta < 1$). Beyond weak selection, increasing selection has a positive effect under perfect payoff memory but a negative effect under limited payoff memory.

22

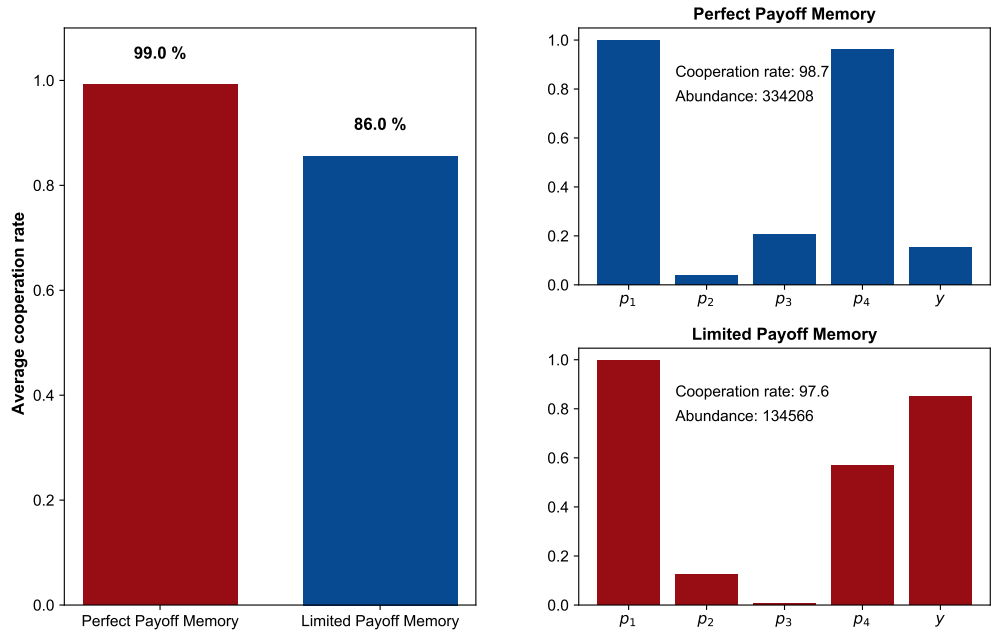

**Supplementary Figure 6: Evolution of memory-one strategies when the benefit is high.** We consider a similar setup as in Supplementary Fig. 5, but now using a larger benefit value $b=10$.

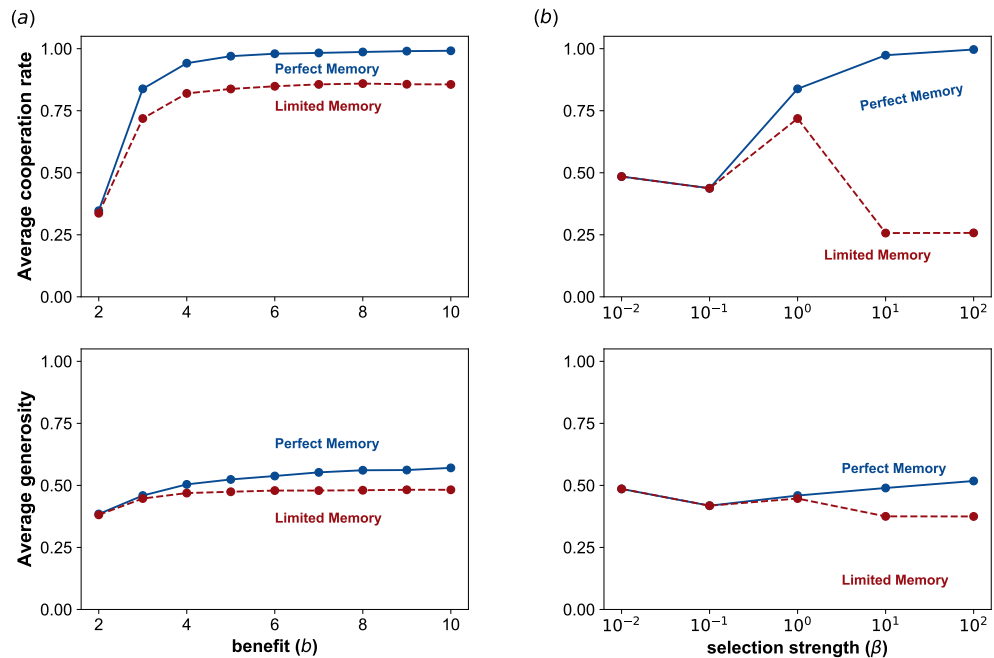

**Supplementary Figure 7: Evolution of direct reciprocity for different parameter values for memory-one strategies.** To further explore the robustness of our results in the case of memory-one strategies, we run simulations for different benefit values (left panels, A) and for different selection strengths (right panels, B). These simulations suggest that perfect payoff memory consistently leads to more cooperation and more generosity. Unless explicitly varied, the parameters of the simulation are $N=100$, $b=3$, $c=1$, $\beta=1$, $\delta=0.999$. Simulations are run for $T=5\times 10^7$ time steps for each parameter combination.



\end{document}